\documentclass[11pt,a4paper,american,floatfix,superscriptaddress,onecolumn,amsmath,amssymb,reprint]{revtex4}
\usepackage{ae,aecompl}
\usepackage[T1]{fontenc}
\usepackage[latin9]{inputenc}
\usepackage{color}
\usepackage{babel}
\usepackage{units}
\usepackage{amsmath}
\usepackage{amssymb}
\usepackage{graphicx}
\usepackage{wasysym}
\usepackage[unicode=true,
 bookmarks=true,bookmarksnumbered=false,bookmarksopen=false,
 breaklinks=false,pdfborder={0 0 0},pdfborderstyle={},backref=false,colorlinks=true]
 {hyperref}
\hypersetup{
 pdfauthor={Dalson Almeida}}

\makeatletter

\pdfpageheight\paperheight
\pdfpagewidth\paperwidth

\@ifundefined{textcolor}{}
{%
 \definecolor{BLACK}{gray}{0}
 \definecolor{WHITE}{gray}{1}
 \definecolor{RED}{rgb}{1,0,0}
 \definecolor{GREEN}{rgb}{0,1,0}
 \definecolor{BLUE}{rgb}{0,0,1}
 \definecolor{CYAN}{cmyk}{1,0,0,0}
 \definecolor{MAGENTA}{cmyk}{0,1,0,0}
 \definecolor{YELLOW}{cmyk}{0,0,1,0}
}

\usepackage{bbm}
\usepackage[latin9]{inputenc}

\makeatother

\begin{document}
\title{Entanglement entropy in multi-leg Kitaev ladders with interface defects}
\author{Dalson Eloy Almeida}
\address{Centro Federal de Educa{\c c}{\~a}o Tecnol{\'o}gica de Minas Gerais, Campus V,
R. {\'A}lvares de Azevedo, 400, Divin{\'o}polis, MG, 35503-822 Brazil}
\begin{abstract}
The entanglement of different parts of a quantum system is expected
to be proportional to the common interface area. Therefore alterations
across the interface will lead to changes on the behavior of entanglement
entropy. In this work, the effects of bond defects at the boundaries
of Kitaev ladders are considered. We find a logarithmic scaling for
the ground state entanglement entropy between the two pieces. The
prefactor of the logarithm (effective central charge) varies continuously
with the defect strength. The energy dispersion is also obtained and
sharp features in the von Neumann entanglement entropy are observed
when bands cross. Phase diagrams for homogeneous Kitaev hamiltonians
with nonzero superconducting paring potential are presented. They
show that for chains/legs that are connected to one another through
inter-leg hopping, when certain parameters are fine-tuned, the phase
transition lines correspond to either single or double gapless modes
dispersion. Moreover, even when the defect is turned on, the effective
central charge for ladders with two gapless modes is exactly twice
the one for ladders with a single gapless mode. On the other hand,
in the absence of superconductivity, we can tune the parameters to
obtain homogeneous systems whose number of gapless modes is up to
the number of legs of the ladder. Additionally, in this situation,
the presence of the bond defect makes the effective central charge
becomes smaller than the number of gapless modes times the effective
central charge of hamiltonians with one gapless mode. Furthermore,
the relationship between the cases with and without superconductivity
is presented.
\end{abstract}
\maketitle

\section{Introduction}

In recent times the theoretical understanding of quantum entanglement
has grown significantly in the condensed matter physics community.
Moreover, the entanglement property of free-fermion systems has emerged
as a key tool of study whenever quantum phase transitions are present.\cite{Vidal_Latorre,Vidal_Latorre2,CALABRESE_CARDY,CALABRESE_CARDY2,AFFLECK2,KOREPIN}
The phase diagram of several one-dimensional models have been analyzed/determined
using such measurements. It is well known that, for homogeneous non-critical
systems, the ground state entanglement entropy $S$ approaches a constant
for large size $L$ of the subsystem, and for critical ones, it logarithmic
divergence with $L$. Many other studies have also investigated non-homogeneous
systems, and various different cases have been explored. For example,
the interface of one-dimensional systems with short-range interactions
reduces to a point, therefore, either a bond or a site single defect
is expected to have a noticeable effect in the correlation between
the subsystems.\cite{Igl_i_2007,Peschel_INTERFACE} Other inhomogeneities
have also been investigated in the past, e.g., coupled impurity \cite{Eisler_impurity},
random and exponentially decaying couplings \cite{Laflorencie,Vitagliano_2010},
and so on.

One of the simplest models of free spinless fermions that one can
consider is the Kitaev chain.\cite{Kitaev_2001} This is a canonical
system that admits a topological phase transition. Beyond this one-dimensional
model, in recent years a generalization of this toy model, involving
more than one chain, have been considered.\cite{Potter_KitaevLadder,Maiellaro,RituNehra}
In this case, the simultaneous presence of both the ladder geometry
and the superconducting term lead to a richer phase diagram. 

When it comes to homogeneous quasi-one-dimensional systems, a couple
of studies have also considered their bipartite entanglement entropy
behavior in the past. For example, the R{\'e}nyi entropy of the ground
state and low-lying excited states of a quadratic spinless fermions
hamiltonian with two-leg geometry was considered in Ref. \cite{Dalson2012}.
The scaling behavior of the ground state entanglement entropy of critical
$N$-leg free fermions ladders is also known, moreover, interacting
systems such as Heisenberg ladders and Ising ladder \cite{Xavier_2014}
have been considered. In this work, we consider inhomogeneities in
$N$-leg ladders of free spinless fermions quadratic hamiltonians.
The system is consisted of two equal subsystem halves and we study
the effect of defects across the interface on effective central charge
(ECC).

A further feature in our work is an investigation of the entanglement
in bi-quadratic hamiltonians that we call \textit{Kitaev $N$-ladder}.
We show that the homogeneous Kitaev hamiltonian undergoes a phase
transition when certain parameters are fine-tuned. Moreover, in the
presence of a non-zero superconducting gap, for connected legs the
critical systems can only have one or two gapless modes on their energy
dispersion. This is quite different from the case in absence of superconducting
pairing potential (quadratic hamiltonians) where the number of gapless
modes ($n_{\mathrm{GL}}$) is up to the number of legs \cite{Xavier_2014}.
Finally, a study of inhomogeneities across the interface is also conducted.

The paper is organized as follows. In Sec. \ref{sec:The-model} we
present the model and its energy dispersion. In Sec. \ref{sec:Gapless-points}
we analyze the phase transition lines for the case with a finite superconducting
gap. Our results for the entanglement in the many-body ground eigenstates
of bi-quadratic hamiltonians are presented in Sec. \ref{sec:Entanglement-entropy}.
Finally, we offer some concluding remarks in Sec. \ref{sec:Conclusions}.

\section{\label{sec:The-model}The model}

\begin{figure}
\begin{centering}
\includegraphics[width=1\textwidth]{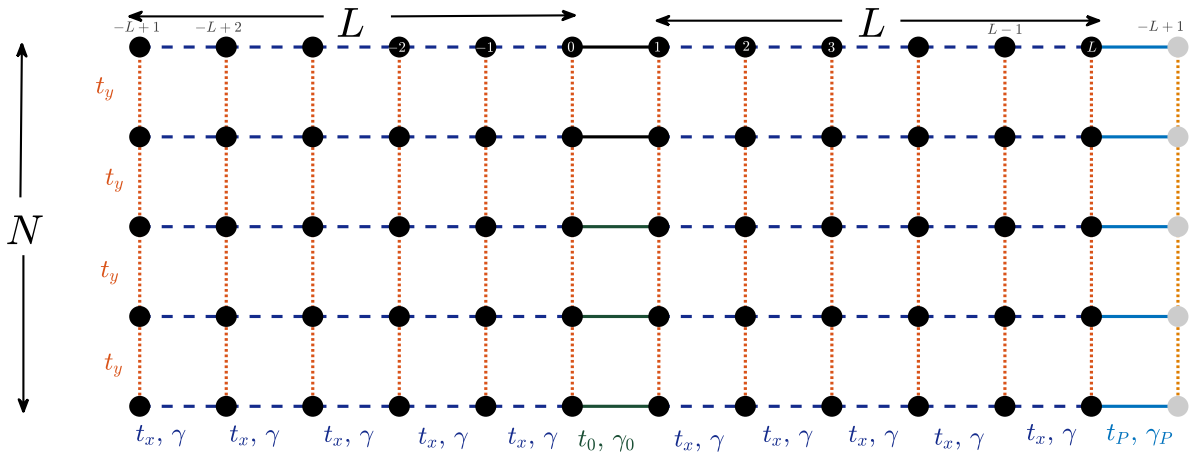}
\par\end{centering}
\caption{Schematic representations of the $N$-leg Kitaev ladder. The system
is composed by $2L$ rungs (vertical), each one of them have $N$
sites. The hopping mechanism (dotted orange line) along these rungs
$t_{y}$ is homogeneous through the ladder. The last rung showed on
the right is a replica of the first one on the left and represents
a possible boundary condition --- the number of the rung is presented
on the top leg. Along the legs (horizontal) we divide the system into
two subsystem with $L$ rungs each. In each subsystem, the hopping
amplitude and the superconducting pairing potential along the legs
(dashed blue lines) are $t_{x}$ and $\gamma$, respectively. On the
other hand, these parameters are equal to $(t_{0},\gamma_{0})$ between
the rungs $n=0$ and $n=1$, and $(t_{P},\gamma_{P})$ between the
rungs $n=L$ and $n=-L+1$. The connection between this two last rung
represents a possible boundary condition between the two subsystems.\label{FIG:interface_retangular}}
\end{figure}

We are going to consider $N$ Kitaev chains \cite{Kitaev_2001}, with
$2L$ sites each, connected to one another through inter-leg hopping
--- a schematic diagram is given in Fig. \ref{FIG:interface_retangular}.
 The hamiltonian of this rectangular Kitaev ladder is given by
\begin{equation}
H=\sum_{n=-L+1}^{L}\left\{ \sum_{\ell=1}^{N}\left[\left(t_{x\,(n)}c_{n,\ell}^{\dagger}c_{n+1,\ell}+\gamma_{n}c_{n,\ell}c_{n+1,\ell}+\mathrm{h.c.}\right)-\mu c_{n,\ell}^{\dagger}c_{n,\ell}\right]+\sum_{\ell=1}^{N-1}\left(t_{y}c_{n,\ell}^{\dagger}c_{n,\ell+1}+\mathrm{h.c.}\right)\right\} \label{eq:H-retangular}
\end{equation}
where $t_{x\,(n)}$, $\gamma_{n}$, and $\mu$ are the nearest neighbor
hopping amplitude along the legs, superconducting pairing potential,
and chemical potential, respectively; $t_{y}$ is the hopping strength
along the rungs, and $c_{n,\ell}$ and $c_{n,\ell}^{\dagger}$ are
the second-quantized creation and annihilation operators at rung $n$
and leg $\ell=1,2,\cdots,N$ . Furthermore, we have the following
boundary condition $c_{n+L,\ell}=c_{n-L,\ell}$ along the legs and
their hoppings and superconducting gaps are given by
\begin{equation}
t_{x\,(n)}=\begin{cases}
t_{0} & n=0\\
t_{P} & n=L\\
t_{x} & \mathrm{otherwise}
\end{cases}\,\,\,\,\mathrm{and}\,\,\,\,\gamma_{n}=\begin{cases}
\gamma_{0} & n=0\\
\gamma_{P} & n=L\\
\gamma & \mathrm{otherwise}
\end{cases}.\label{eq:hoppings}
\end{equation}

For simplicity we are going to only consider open boundary conditions
(OBC) along the rungs. On the other hand, along the chains/legs different
boundary conditions will be imposed. For example, for $t_{0}=t_{x}$,
$t_{P}=0$, $\gamma_{0}=\gamma$, and $\gamma_{P}=0$, the ladder
is homogeneous and has OBC along the chains. Another option is $t_{x}=t_{0}=t_{P}$
and $\gamma=\gamma_{0}=\gamma_{P}$, in this case the system is again
homogeneous but it has periodic boundary conditions (PBC). We will
also investigate homogeneous ladder with anti-periodic boundary conditions
(anti-PBC), i.e., $t_{x}=t_{0}=-t_{P}$ and $\gamma=\gamma_{0}=-\gamma_{P}$.
It is worth mentioning that regardless of the boundary condition that
is set al.ong the chains, in the thermodynamic limit, $L\to\infty$,
the same quantitative critical behavior is expected to be observed.

In general, we can study the hamiltonian \ref{eq:H-retangular} using
a formalism developed by Lieb \textit{et al.} in 1961.\cite{LIEB_ANN}
They investigated systems of $\mathcal{N}$ fermions governed by bi-quadratic
hamiltonians, so, the Hilbert space size is $2^{\mathcal{N}}$. They
showed that the diagonalization of the hamiltonian can be reduced
to the solution of a eigenvalue equation of an $\mathcal{N}$ by $\mathcal{N}$
matrix. In general, in the presence of inhomogeneities, we determine
numerically the eigenvalues and eigenvectors of this matrix in order
to obtain the full energy spectrum.

On the other hand, for the homogeneous cases, that we have mentioned
before, the hamiltonian {[}Eq. (\ref{eq:H-retangular}){]} can be
straight forward diagonalized by standard Fourier transformations.
This leads to a dispersion that consists of $2N$ bands with energies
\begin{equation}
\Lambda_{k,q}=\pm\sqrt{\Delta_{k}^{2}+\epsilon_{k,q}^{2}}\label{eq:dispersion}
\end{equation}
with 

\begin{equation}
\epsilon_{k,q}=2\,t_{x}\cos k+2\,t_{y}\cos q-\mu,\label{eq:epsilon}
\end{equation}
\begin{equation}
\Delta_{k}=2\gamma\sin k,\label{eq:Delta}
\end{equation}
and 
\begin{equation}
q=\frac{\pi}{N+1},\frac{2\pi}{N+1},\cdots,\frac{N\pi}{N+1}.\label{eq:q-valores}
\end{equation}
$\{k\}$ is a set of $2L$ numbers that depends on the boundary condition
along the legs. More details of this calculation calculation can be
found, for example, in Ref. \cite{Nleg_Wang}, where Wang \textit{et
al.} have determined the phase diagrams and the Majorana zero modes
of multi-leg Kitaev ladders.

If PBC are imposed along the rungs the system will have a tube geometry
instead of the stripe geometry considered. For the homogeneous case,
that simply means that the momentum set of Eq. (\ref{eq:q-valores})
becomes

\begin{equation}
q=0,\frac{2\pi}{N},\frac{4\pi}{N},\cdots,\frac{2(N-1)\pi}{N},\label{eq:a-pbc}
\end{equation}
for $N>1$, and therefore the phase diagrams that we are going to present
in the next section would have to be re-obtained. On the other hand,
the ECC general behavior \textit{will not be changed} by this boundary
condition.

\section{\label{sec:Gapless-points}Gapless points}

\begin{figure}
\begin{centering}
\includegraphics[width=0.33\textwidth]{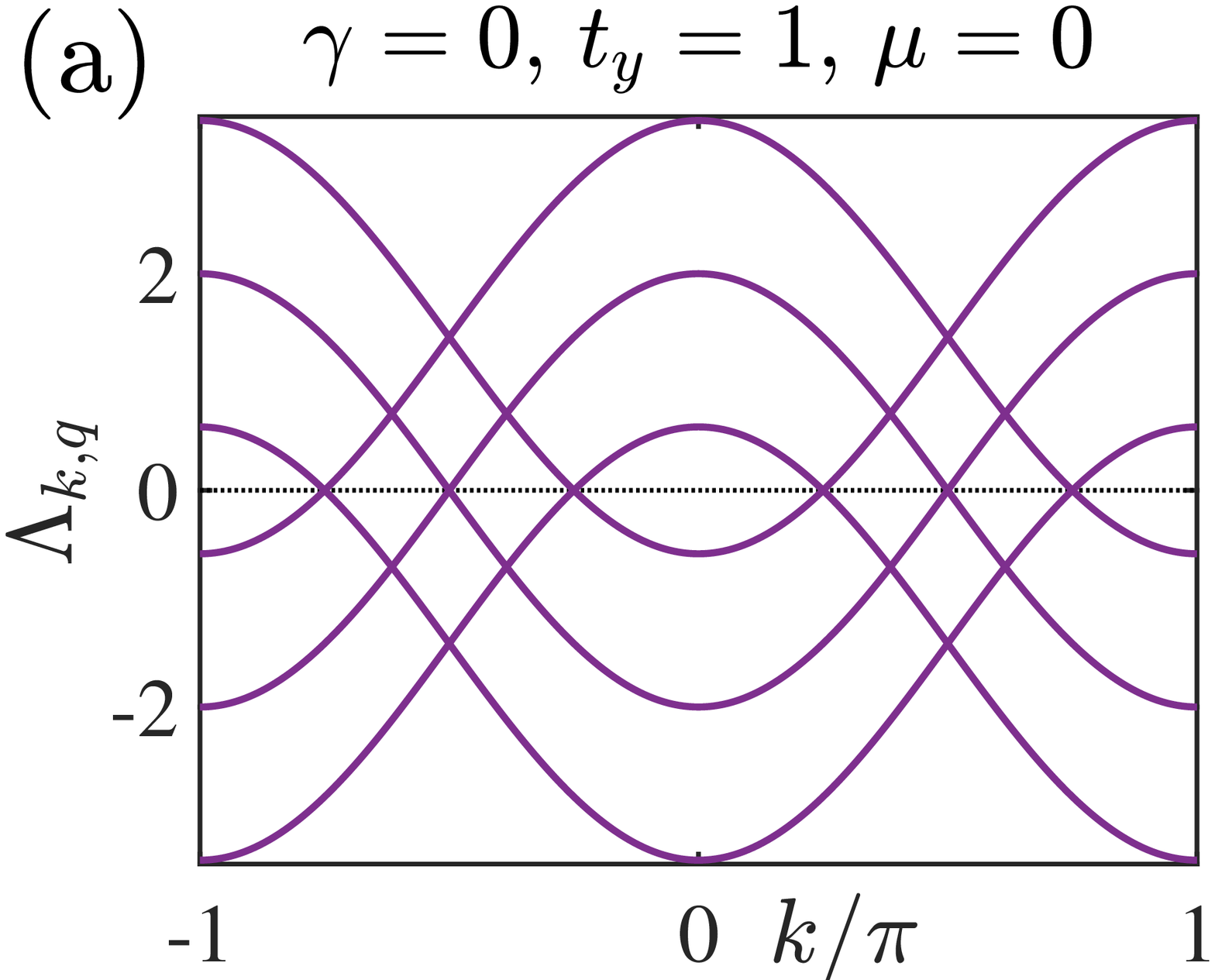}\includegraphics[width=0.33\textwidth]{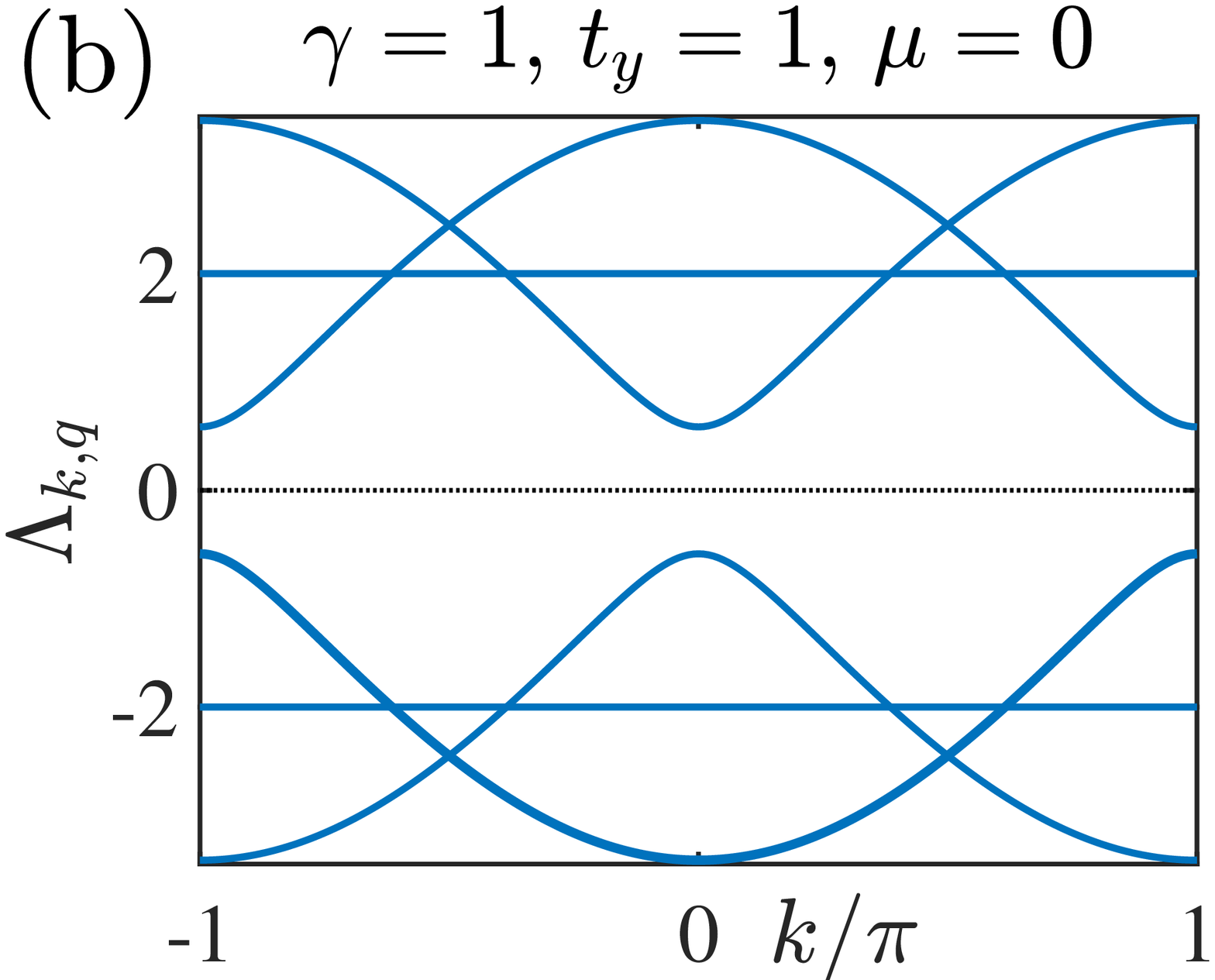}\includegraphics[width=0.33\textwidth]{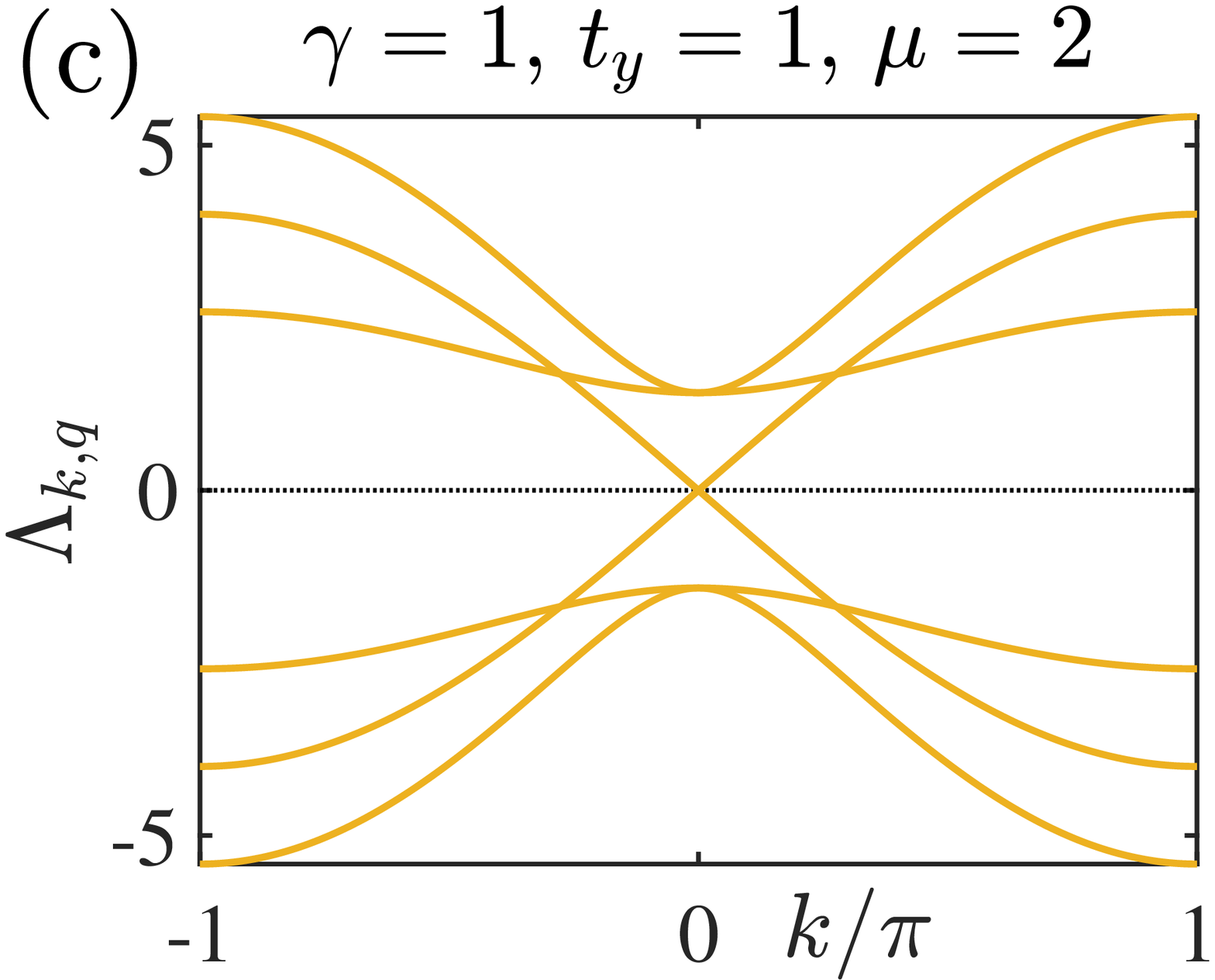}
\par\end{centering}
\vspace{3mm}
\begin{centering}
\includegraphics[width=0.33\textwidth]{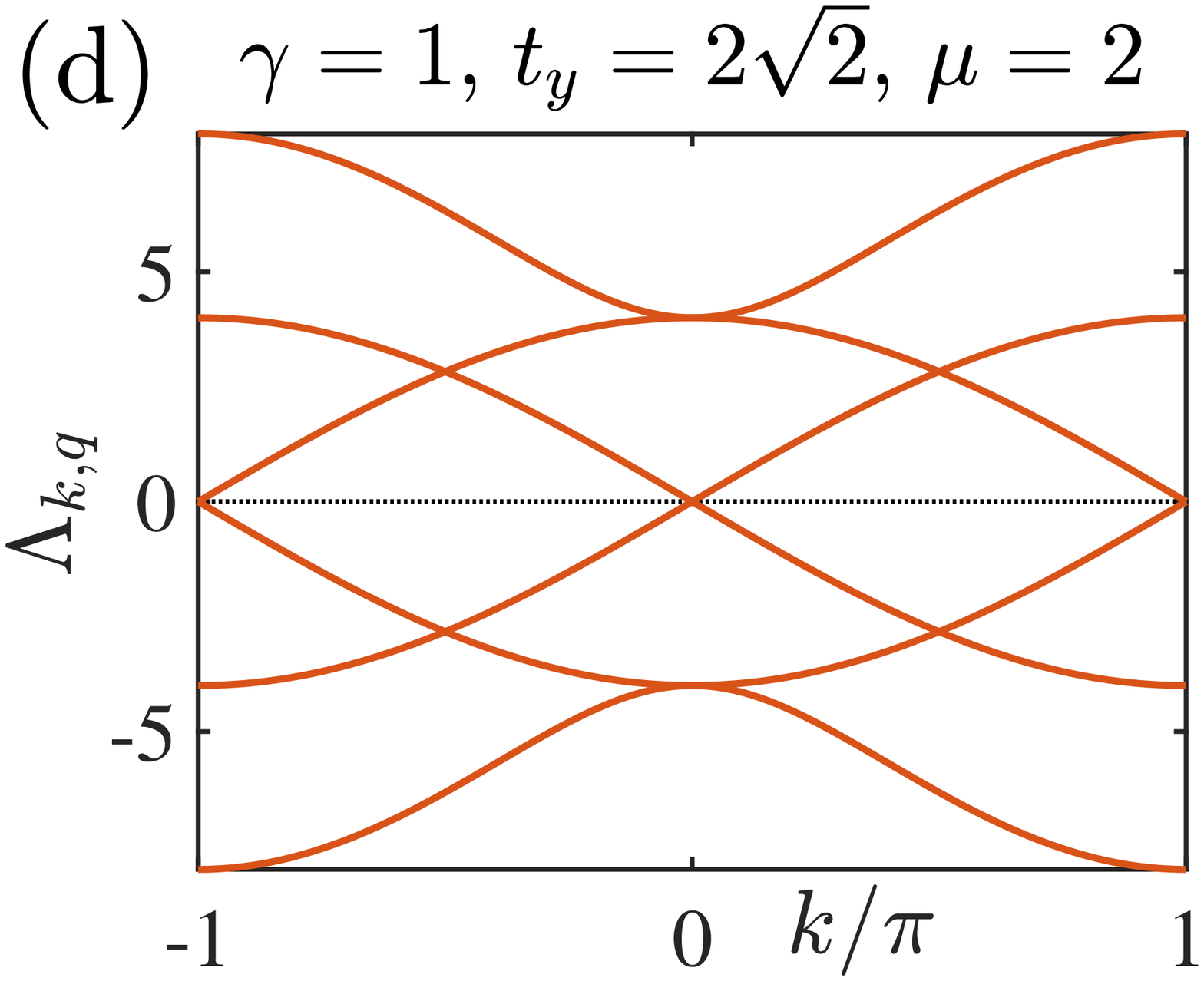}\includegraphics[width=0.33\textwidth]{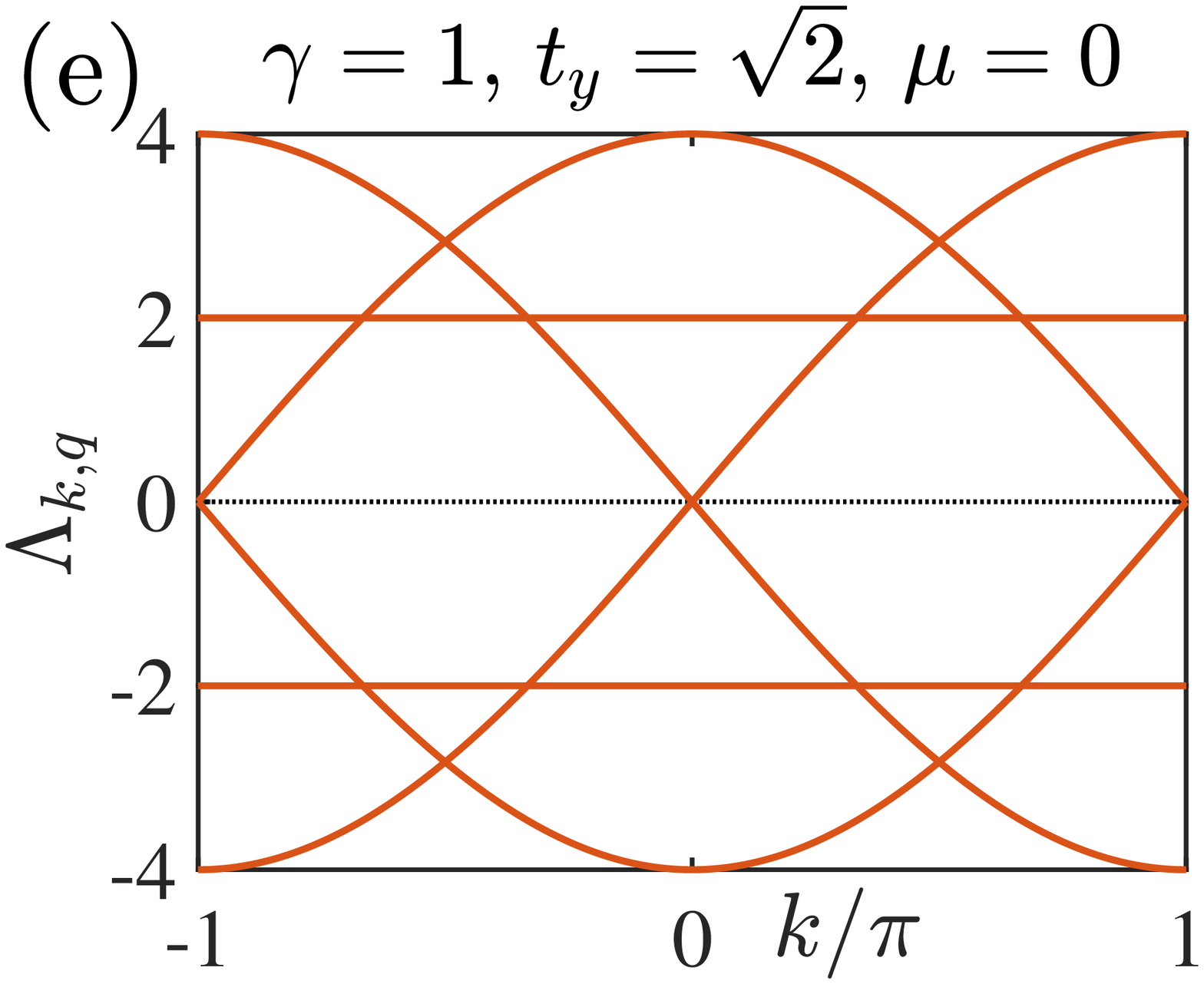}\includegraphics[width=0.33\textwidth]{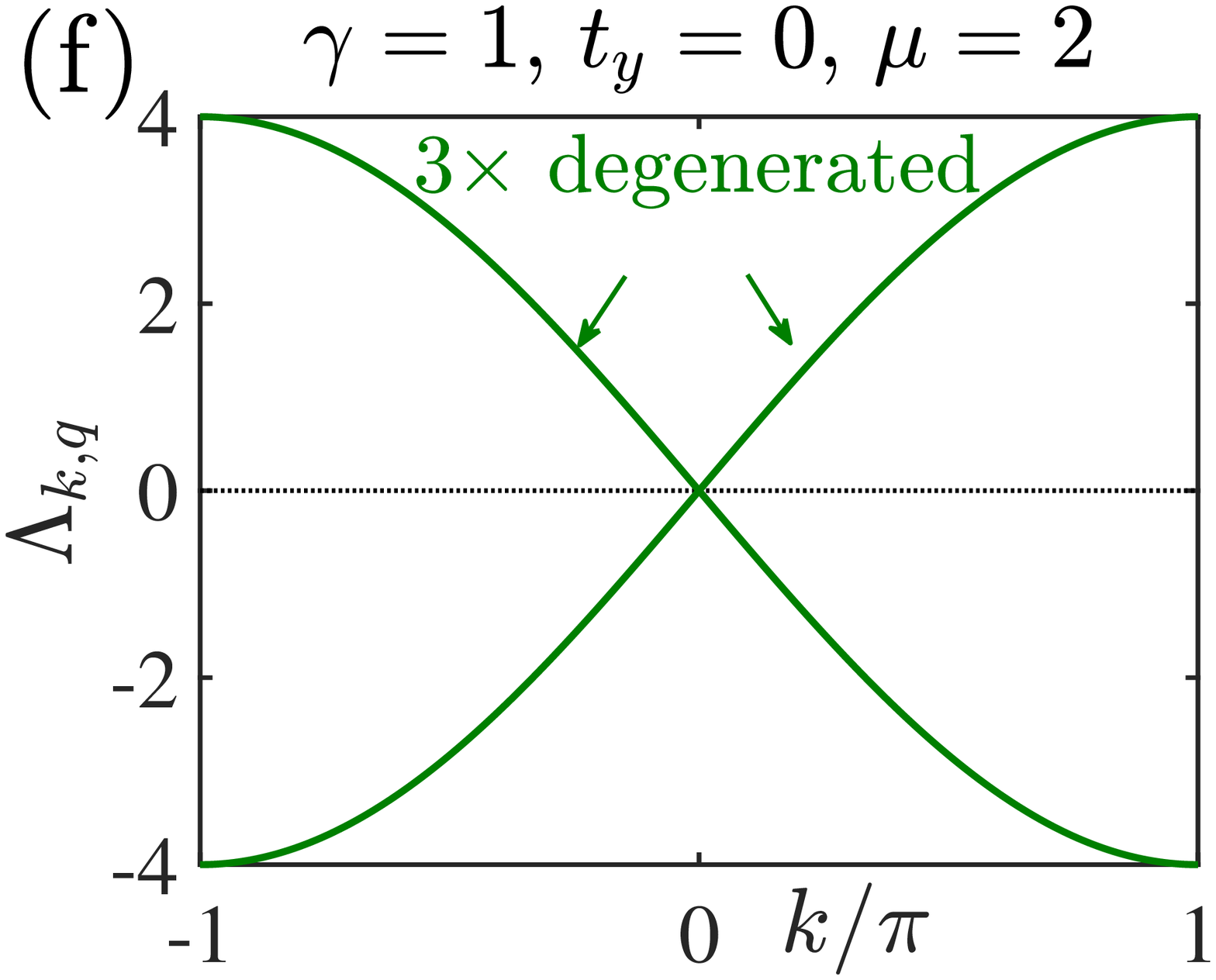}
\par\end{centering}
\caption{$N=3$ spectra --- (a) three gapless excitations on a quadratic hamiltonian
in the absence for superconductivity gap $\gamma=0$; The other panels
show cases with $\gamma=1$, namely: (b) gapped excitations; (c) excitations
with one gapless mode at $k=0$; (d) and (e) two gapless modes at
both $k=0$ and $|k|=\pi$; (f) each line is three times degenerated,
therefore there are three gapless modes $k=0$. \label{fig:dispersao_N=00003D3}}
\end{figure}
In this section we determine the phase diagram for the \textit{homogeneous}
Kitaev ladders. In the thermodynamic limit ($L\to\infty$) the excitations
are going to be gapless if, and only if, $\Lambda_{k,q}\to0$. Additionally,
the condition $\Lambda_{k,q}=0$ is reached only if $\epsilon_{k,q}=0$
and $\Delta_{k}=0$ \textit{simultaneously}. The second requirement
is full filled with either $\gamma=0$ or $\sin k=0$ --- see Eq.
\ref{eq:Delta}.\textcolor{red}{{} }As an example of the well known
case without the superconducting pairing potential, i.e., $\gamma=0$,
we show, in Fig. \ref{fig:dispersao_N=00003D3}(a), the energy dispersions
for a system with PBC, $t_{y}=t_{x}=t_{0}=t_{P}=1$, $N=3$ legs,
and in the absence of chemical potential ($\mu=0$) . Notice that
there are exactly three gapless modes ($n_{\mathrm{GL}}=3$) for $k>0$.
In general, for an arbitrary $N$, if we tune $t_{y}=t_{x}=t_{0}=t_{P}=1$
and $\mu=0$ the dispersion profile have $n_{\mathrm{GL}}=N$.

Let us focus our attention on the case where $\sin k=0$. This situation
is satisfied by either $k=0$ or $|k|=\pi$. Substituting this values
for $k$ into Eq. (\ref{eq:epsilon}) and also requiring that $\epsilon_{k,q}=0$
lead to the following condition of the critical chemical potential
\begin{equation}
\mu_{\mathrm{cr}}=2\,(t_{y}\cos q\pm t_{x}).\label{eq:mu_cr}
\end{equation}
In other words, if the chemical potential is set to be $\mu=\mu_{\mathrm{cr}}$
the ladder will have gapless excitations in the thermodynamic limit.
It is worth mentioning that the above critical lines do not depend
on the superconducting gap $\gamma$. Hereafter we will \textit{fix
a energy scale by setting} $t_{x}=1$. 

In the following subsections we will discuss the behavior of the critical
lines for the cases $N=1,2,3$, and $4$. For a given $N$ there are
$N$ different values for $q$ {[}given by Eq. (\ref{eq:q-valores}){]},
thus, there will be $2N$ critical lines in a $t_{y}-\mu_{\mathrm{cr}}$
plane phase diagram. Also, we will only show results for $t_{y}\geqslant0$
because $\mu_{\mathrm{cr}}$ is symmetric with respect to mirror reflection
$t_{y}\leftrightarrow-t_{y}$.
\begin{figure}
\begin{centering}
\includegraphics[width=0.5\textwidth]{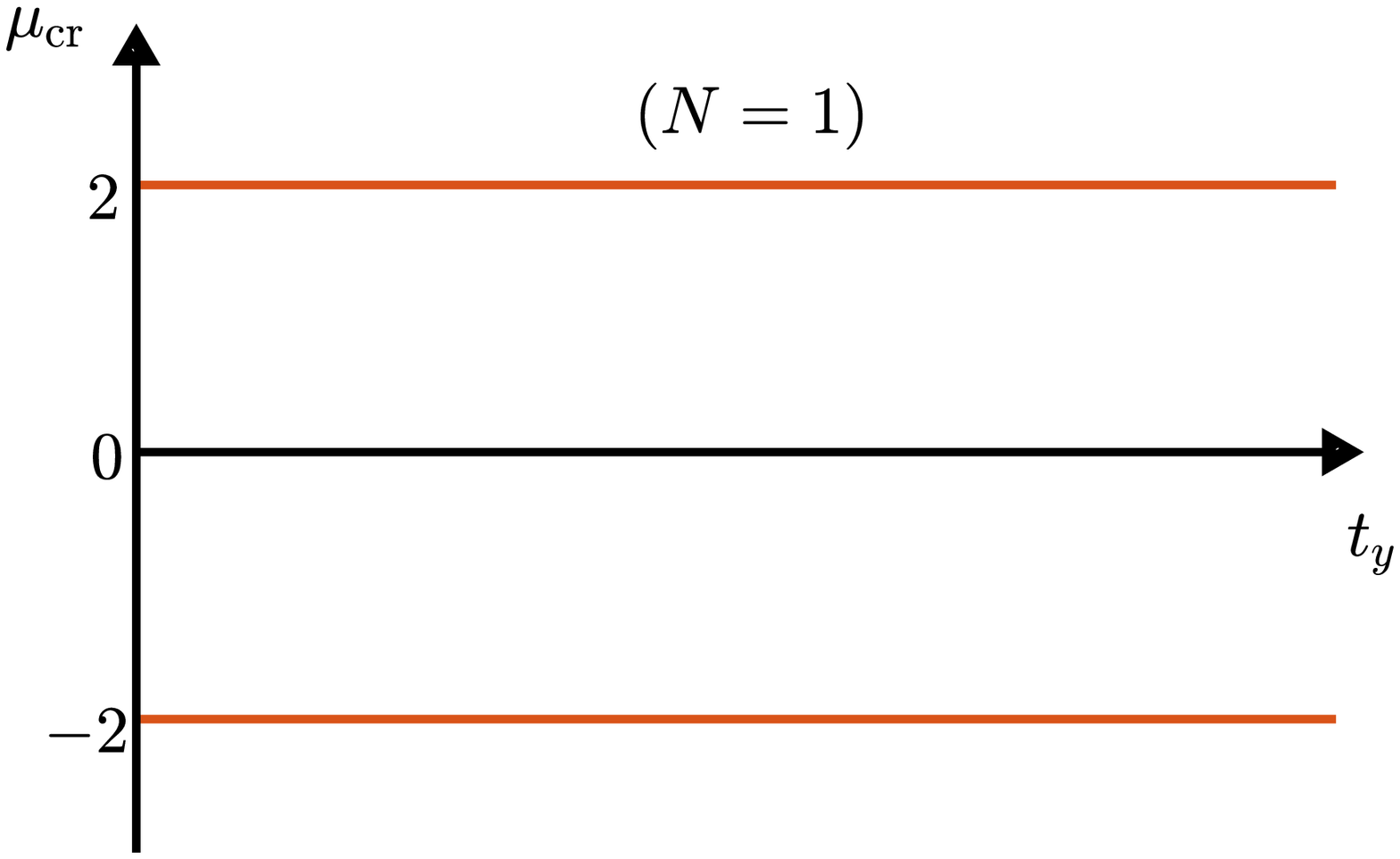}\includegraphics[width=0.5\textwidth]{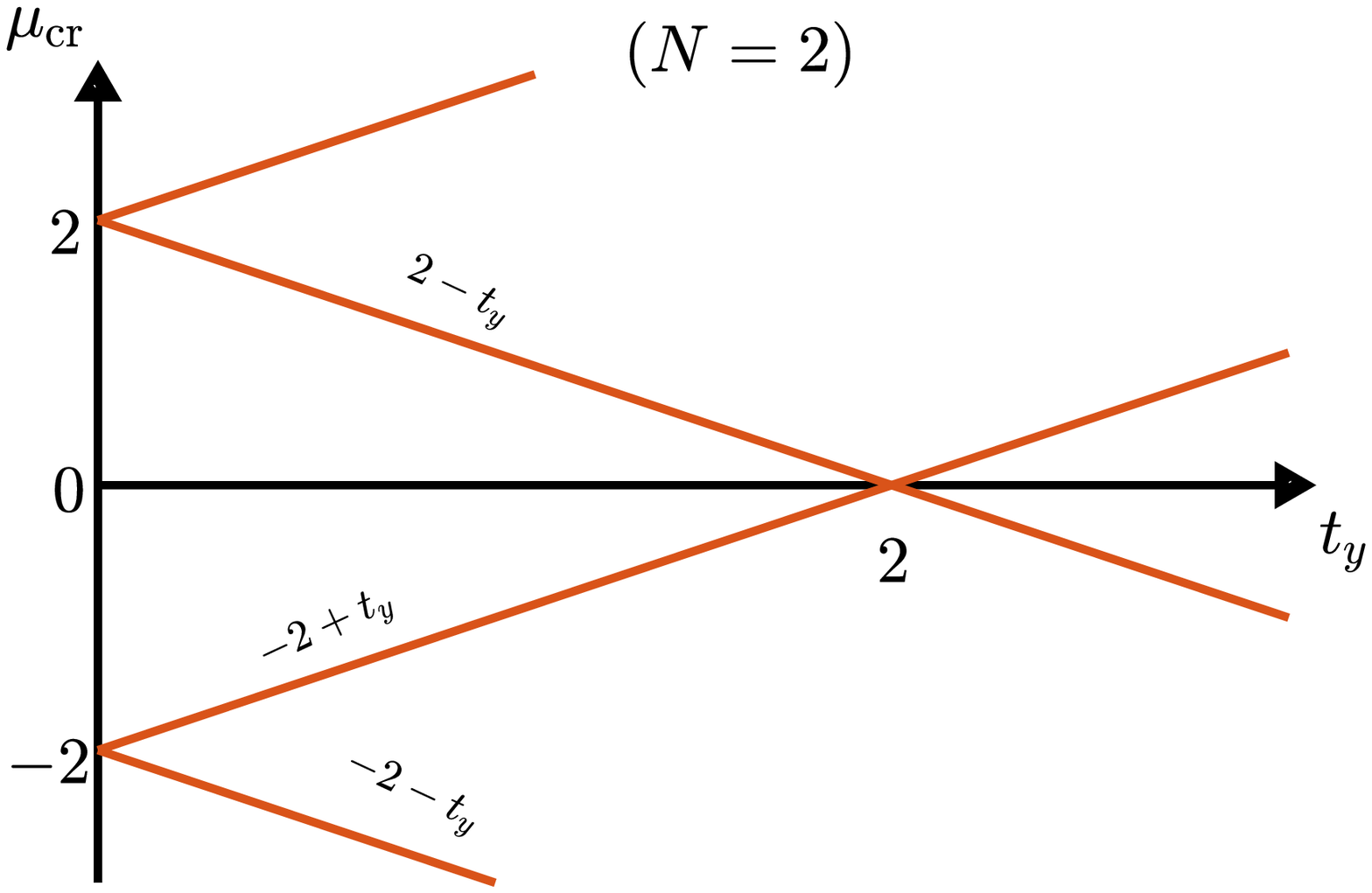}
\par\end{centering}
\begin{centering}
\includegraphics[width=0.5\textwidth]{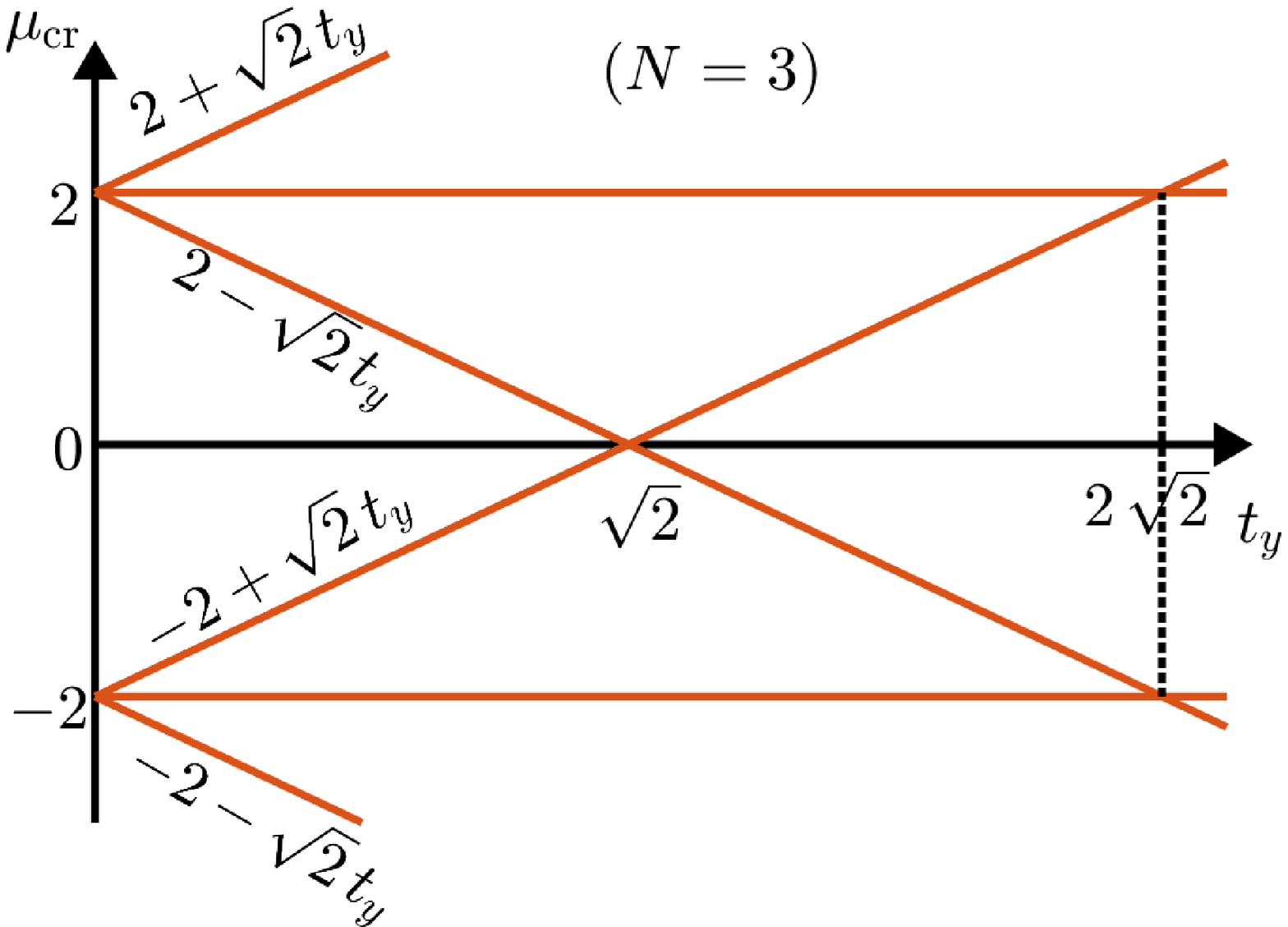}\includegraphics[width=0.5\textwidth]{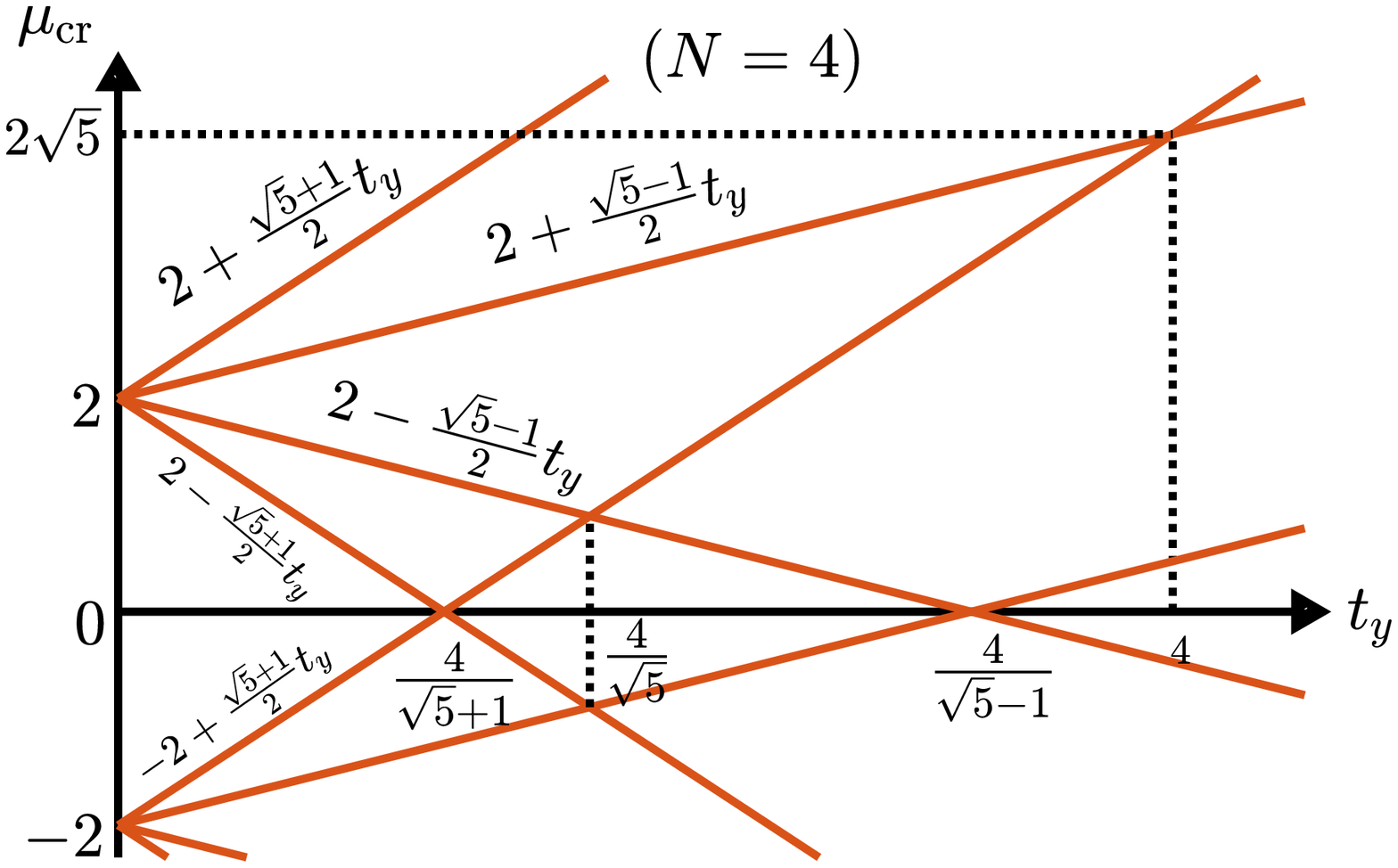}
\par\end{centering}
\caption{Critical chemical potential as functions of $t_{y}$ (we have set
$t_{x}=1$ as the energy scale) for $N=1,2,3$, and $4$. We refrain
from showing the behavior of $\mu_{\mathrm{cr}}$ for $t_{y}<0$ because
it can be obtained by reflection over the $\mu_{\mathrm{cr}}$-axis.
These lines are independent of the finite superconducting gap $\gamma$
value.\label{fig:critical-mu}}
\end{figure}

\subsubsection*{$N=1$ --- Kitaev chain}

For the Kitaev chain $(N=1)$, according to Eq. (\ref{eq:q-valores}),
the only $q$ value allowed is $q=\pi/2$. In consequence we have
the well known result that the system is critical if the chemical
potential is set to $\mu=\mu_{\mathrm{cr}}=\pm2$ (see the top left
panel in Fig. \ref{fig:critical-mu}). Moreover, the two critical
lines are parallel to the $t_{y}$-axis and therefore they never intercept
each other, so at these critical lines there is always only one gapless
excitation mode ($n_{\mathrm{GL}}=1$) at either $k=0$ or $|k|=\pi$.

\subsubsection*{$N=2$ --- simple (two-leg) Kitaev ladder}

The two-leg problem has also been discussed for several authors in
the past (see for example Ref. \cite{Nehra_diag,RituNehra,Maiellaro}).
For the sake of clarity we also present the phase diagram of this
ladder in Fig. \ref{fig:critical-mu}. There are four independent
lines given by substituting the vertical momenta $q=\pi/3$ and $q=2\pi/3$
into Eq. (\ref{eq:mu_cr}). These critical chemical potential lines
meet at the points $(t_{y}=0,|\mu_{\mathrm{cr}}|=2)$ and $(|t_{y}|=2,\mu_{\mathrm{cr}}=0)$.
Consequently, the dispersion have two gapless modes at these interception
points, otherwise the critical lines represent situations with $n_{\mathrm{GL}}=1$.
Note that the first case $(t_{y}=0,|\mu_{\mathrm{cr}}|=2)$ represents
two disconnected Kitaev chain, and in this trivial situation the bands
$\Lambda_{k,q=\pi/3}$ and $\Lambda_{k,q=2\pi/3}$ are fully degenerated.
Furthermore, there are no additional interception among the critical
lines. That is because the two increasing/decreasing lines have the
same inclination. 

\subsubsection*{$N=3$ --- three-leg Kitaev ladder}

Let us explore the $N=3$ system with more details. In addition to
the two horizontal lines $\mu_{\mathrm{cr}}=\pm2$, that are present
on the phase diagram for the single chains (and for all odd $N$),
there are four other lines: $\mu_{\mathrm{cr}}=2\pm\sqrt{2}\,t_{y}$
and $\mu_{\mathrm{cr}}=-2\pm\sqrt{2}\,t_{y}$. We exhibit all these
critical lines in Fig. \ref{fig:critical-mu}. Once more, the number
of lines intercepting each other at some point gives the number of
gapless modes in the spectrum. For that reason the excitations have
$n_{\mathrm{GL}}=3$ when $t_{y}=0$ and $|\mu_{\mathrm{cr}}|=2$
(three disconnected chains) and $n_{\mathrm{GL}}=2$ at ($|t_{y}|=\sqrt{2}$,$\mu_{\mathrm{cr}}=0$)
or ($|t_{y}|=2\sqrt{2}$,$|\mu_{\mathrm{cr}}|=2$). Otherwise, the
critical lines represent systems with only one gapless mode excitation.
Again, there is no more interceptions because the two increasing (or
the decreasing) lines are parallel to each other, and that means they
will not have interception points.

In order to understand better the behavior of the dispersions we show
typical band energies as a function of $k$ for five different values
of ($t_{y}$,$\mu$) in Fig. \ref{fig:dispersao_N=00003D3}(b) - (f).
For simplicity we set $\gamma=1$ in all those graphs, but this simplification
does not change the number of gapless mode excitations. In Fig. \ref{fig:dispersao_N=00003D3}(b)
we have used $t_{y}=1$ and $\mu=0$, so the system is non-critical
and the energy spectrum displays a gapped dispersion. In Fig. \ref{fig:dispersao_N=00003D3}(c)
we set $t_{y}=1$ and $\mu=\mu_{\mathrm{cr}}=2$, therefore, as expected
the system is critical with a single gapless mode at $k=0$. In the
panels (d) and (e) we have chosen $t_{y}$ and $\mu=\mu_{\mathrm{cr}}$
such as two critical lines of the phase diagram {[}Fig. \ref{fig:critical-mu}{]}
intercept each other, so there are two gapless modes at both $k=0$
and $|k|=\pi$. Finally, we set $t_{y}=0$ and $\mu=\mu_{\mathrm{cr}}=2$
in Fig. \ref{fig:dispersao_N=00003D3}(f), i.e., we have three disconnected
critical Kitaev chains ($\Lambda_{k=0,q}=0$) and the dispersion is
three times degenerated --- this means that $\Lambda_{k,q=\pi/4}=\Lambda_{k,q=\pi/2}=\Lambda_{k,q=3\pi/4}$
for all $k$.

\subsubsection*{$N=4$ --- four-leg Kitaev ladder}

We now move on to the phase diagram for $N=4$ (presented in the bottom
right panel of Fig. \ref{fig:critical-mu}). Again, we are going to
comment here only the case $t_{y}\geqslant0$ and $\mu_{\mathrm{cr}}\geqslant0$
because the phase diagram is symmetric under reflection about both
$t_{y}$-axis and $\mu$-axis. Once more, the point $(t_{y},\mu_{\mathrm{cr}})=(0,2)$
represents four disconnected chains, as a consequence, the dispersion
is four times degenerated and there are four gapless mode excitations
at $k=0$. The points in the phase diagram where two lines meet ---
leading to $n_{\mathrm{GL}}=2$ --- are given $(t_{y},\mu_{\mathrm{cr}})=(\nicefrac{4}{\sqrt{5}\pm1},0)$,
$(\nicefrac{4}{\sqrt{5}},\nicefrac{2}{\sqrt{5}})$, and $(4,2\sqrt{5})$.
And there is no additional interceptions.

To summarize, for an arbitrary $N$, when the parameters are tuned
to $t_{y}=0$ and $\mu_{\mathrm{cr}}=2$ we are going obtain the trivial
situation with $N$ disconnected chains. So, there will be $N$ degenerated
bands and they all have a gapless excitation modes at $k=0$. Otherwise,
we expect that the number of gapless mode excitations for the critical
ladders will be either one or two. That is because, besides the trivial
situation mentioned above, the interceptions of chemical potential
critical lines in the phase diagram are supposed to happen only between
two of the $2N$ critical at a time.

In the next section we will explore the entanglement entropy between
two parts of the ladder. As expected we can use this quantity to investigate
the critical points that we have discussed here.

\section*{\label{sec:Entanglement-entropy}Entanglement entropy}

An important property of a critical point is the divergence of the
so called correlation length. Moreover, for a system divided into
two subsystems, the subsystems will be bipartite entangled with each
other when there are quantum correlations between them. The von Neumann
many body entanglement entropy $S$ of the ground state of the composite
quantum system is often used to measure the degree of quantum entanglement
between two subsystems. Throughout this section we will \textit{divide
the ladder} {[}given by the hamiltonian (\ref{eq:H-retangular}){]}\textit{
exactly into two equal halves and subsystems will be taken as $n\leqslant0$
and $n>0$}. One can calculated the von Neumann entanglement entropy
$S$ between the two parts of the ladder using the standard correlation
method.\cite{Vidal_Latorre,Vidal_Latorre2,CALABRESE_CARDY,CALABRESE_CARDY2,AFFLECK2,KOREPIN}

\subsection*{S as a function of $\mu$ and $t_{y}$}

\begin{figure}
\begin{centering}
\includegraphics[width=0.5\textwidth]{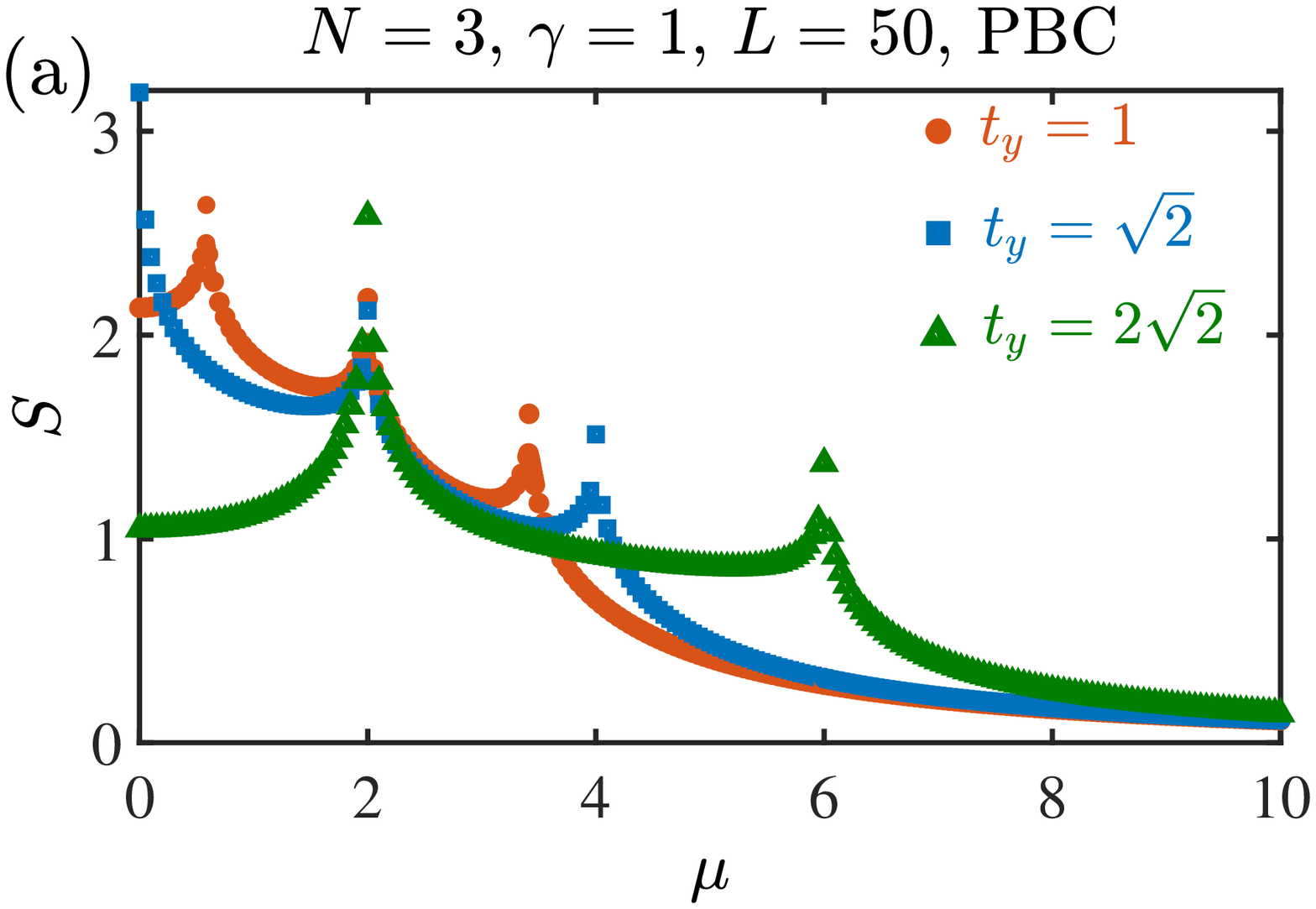}\includegraphics[width=0.5\textwidth]{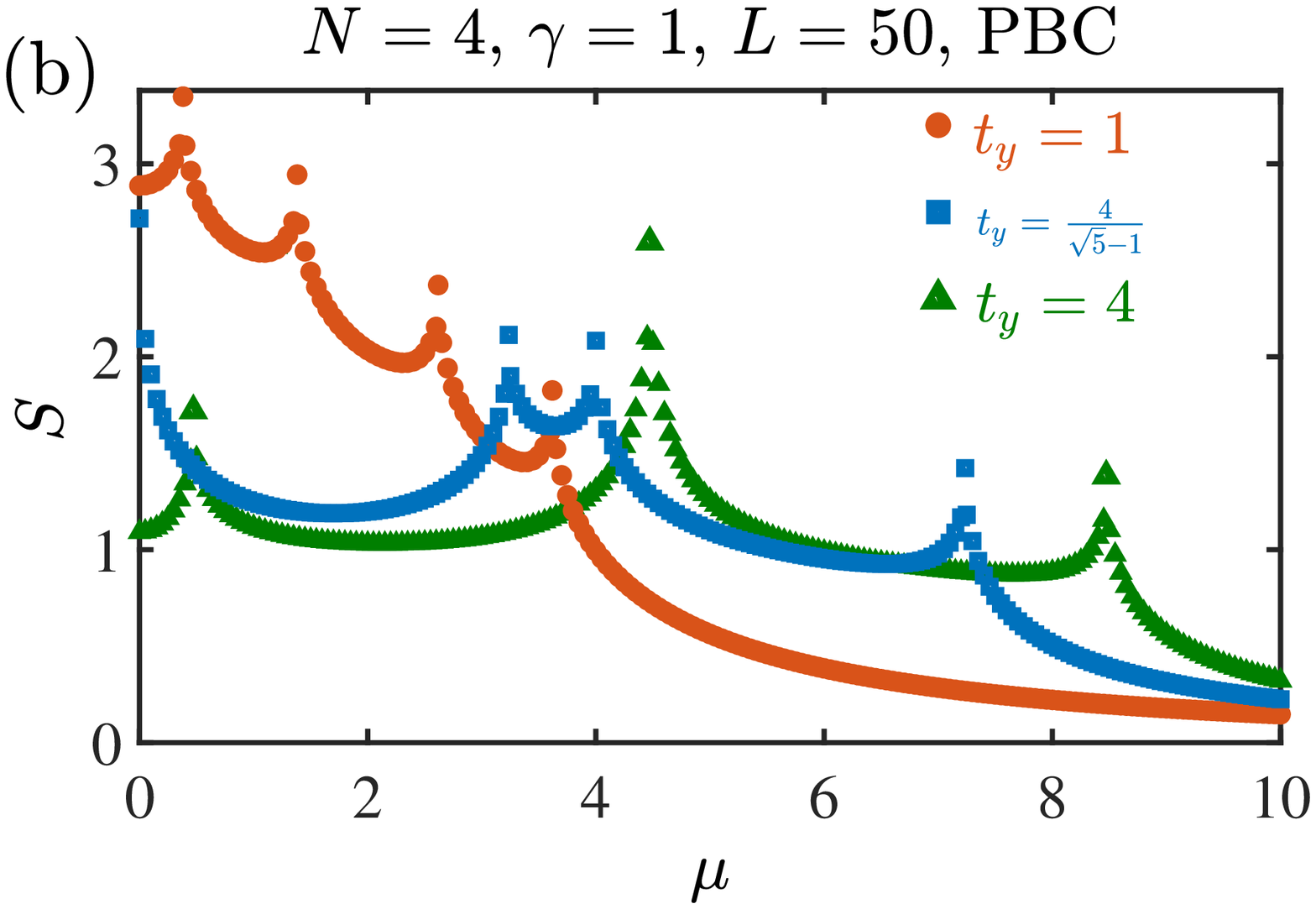}
\par\end{centering}
\vspace{2mm}
\begin{centering}
\includegraphics[width=0.5\textwidth]{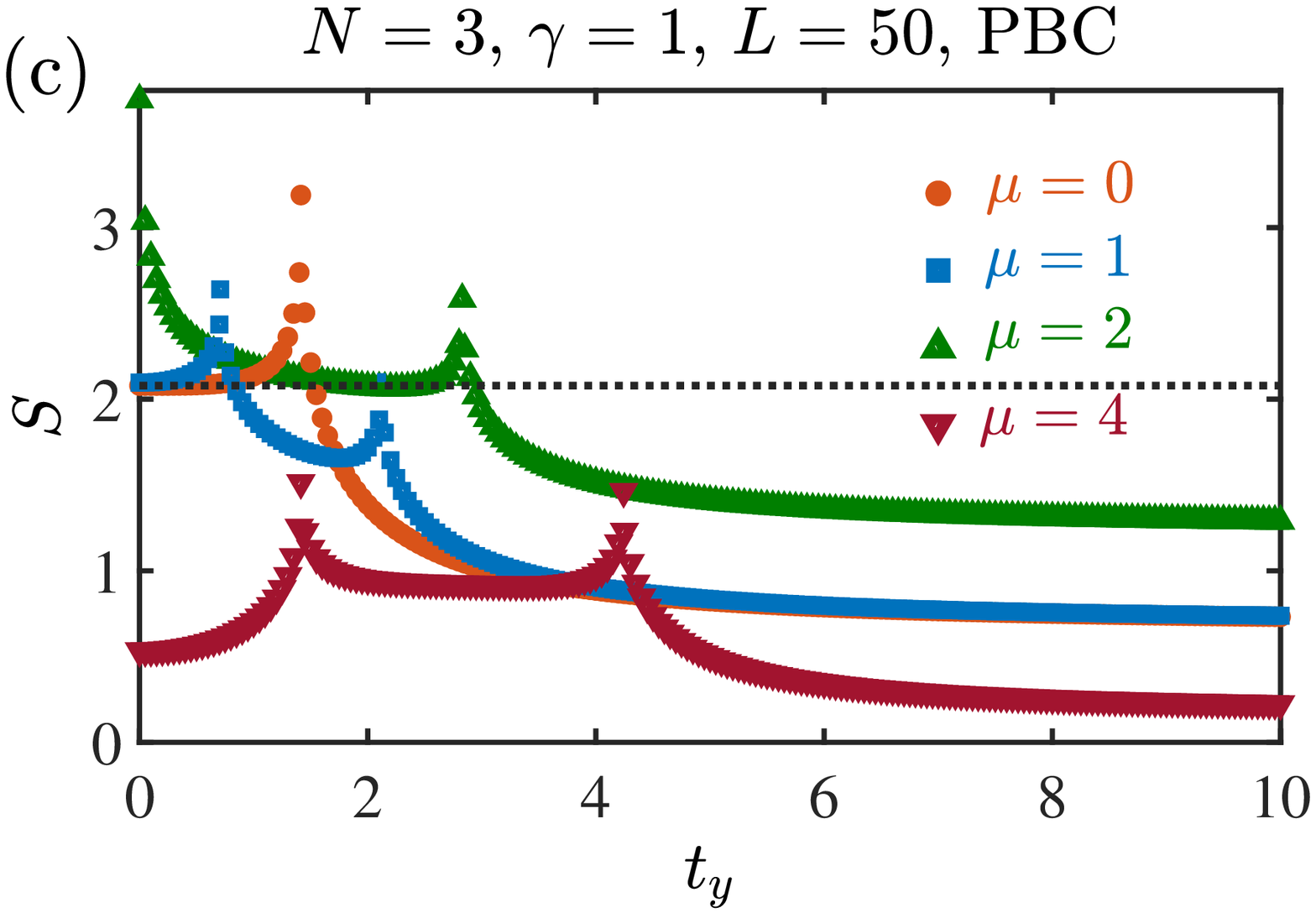}\includegraphics[width=0.5\textwidth]{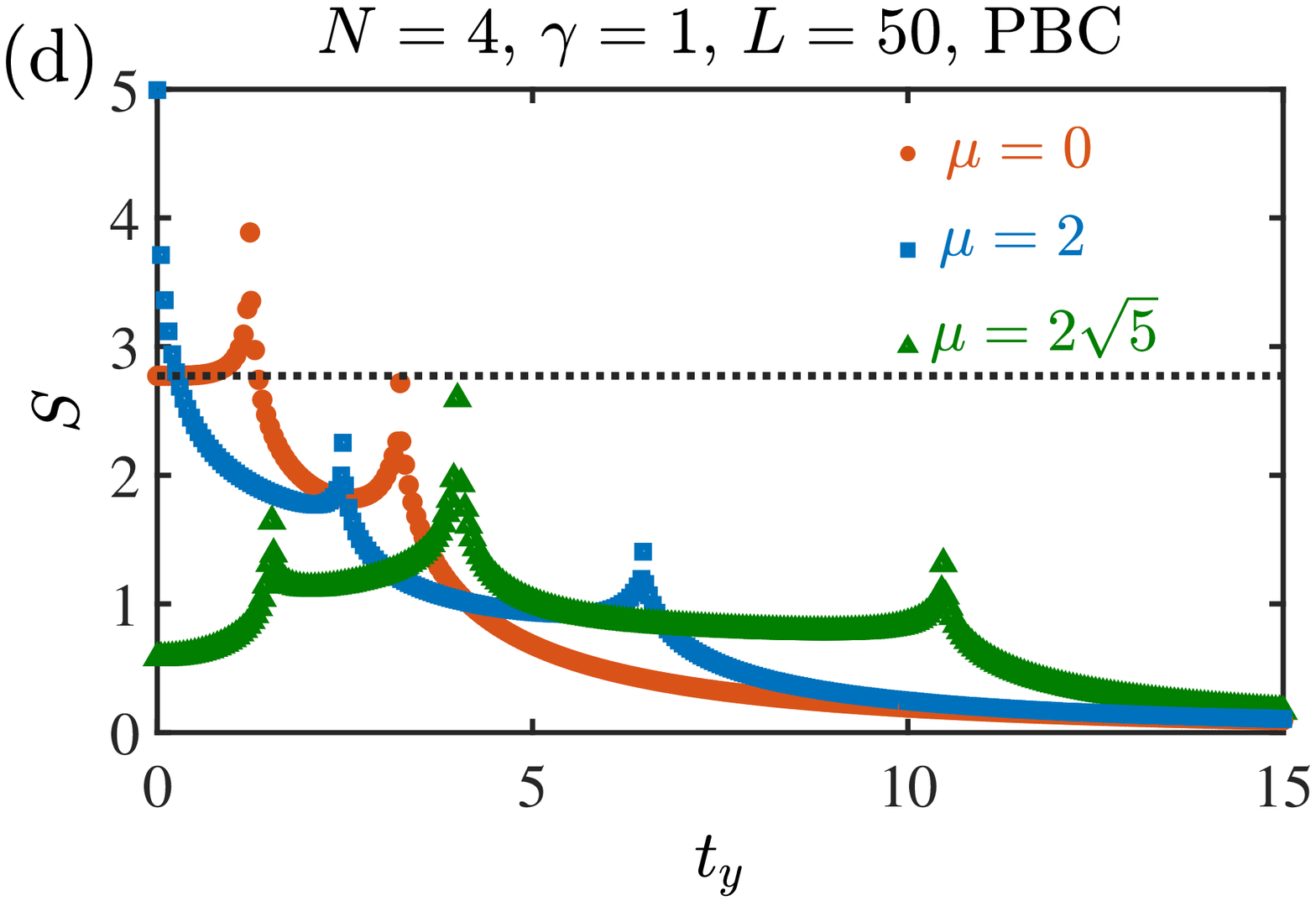}
\par\end{centering}
\caption{Entanglement entropy $S$, under PBC, as a function of: (a) and (b)
the chemical potential $\mu$ for three different values of $t_{y}$;
(c) and (d) the inter-leg hopping $t_{y}$ for fixed values of $\mu$
(the horizontal dotted lines represent $S=N\,\ln2$). \label{fig:S_vs_mu_and_ty}}
\end{figure}
In this subsection, the ground state entanglement entropies between
the left and right sides of \textit{homogeneous} ladders are obtained
as functions of either the hopping term along the rung $t_{y}$ or
the chemical potential $\mu$. We have imposed PBC along chains of
size $2L=100$ and set $\gamma=1$ here in this subsection. Moreover,
we only present data for $t_{y}\geqslant0$ and $\mu\geqslant0$ because
$S$ is also an even function of both $t_{y}$ and $\mu$.

The case $N=1$ (single Kitaev chain) is  analogous to the famous
quantum Ising chain in a critical transverse magnetic field. In fact
this simple case was extensively investigated by several authors in
the past.\footnote{For example, an useful and detailed study of the behavior of the entanglement
entropy of several exactly solvable many body systems can be found
in Refs. \cite{PESCHEL_VIK,PESCHEL_BJP}.} The magnetic field in the Ising chain is equivalent to the chemical
potential here, and it is well know that, for chains with OBC, the
entanglement entropy is going to diverge as $S\sim\frac{1}{6}\ln\,L$
at the critical point. Likewise, Nehra \textit{et al.} \cite{RituNehra}
have recently analyzed the ladder with two legs and they found divergences
on the entanglement entropy at the critical points discussed in the
past Section {[}top right panel in Fig. \ref{fig:critical-mu}{]}.
Therefore, let us focus our attention on the case $N>2$.

In the panels (a) and (b) of Fig. \ref{fig:S_vs_mu_and_ty} we show
$S$ as functions of the chemical potential $\mu$ for $N=3$ and
$N=4$, respectively. In each figure there are three different curves
representing three different hopping $t_{y}$ along the rungs. It
is clear that for several critical values of $\mu$ the systems become
critical --- the signature of this quantum phase transition is the
accentuated peaks that indicate a divergence in $S$ in the thermodynamic
limit $L\to\infty$. Moreover, we notice that for large $\mu$ the
von Neumann entropy tends to approach zero, that is because the ground
state of the system tends to the fully filled state. This is a product
state, so the subsystems will be not entangled. Analogously, if we
consider the case $\mu\to-\infty$ we are going to also obtain $S\to0$
in this other limit. The reason is quite the same, since in this case
the ground state tends to approach the fully empty state, thus quantum
correlation between the left and right part of the ladder disappears.

For the sake of explicitness, we present below the location of the
peaks observed on the entanglement entropy of the three-leg ladder
{[}Fig. \ref{fig:S_vs_mu_and_ty}(a){]}:

- $t_{y}=1$ (orange $\CIRCLE$): $\mu_{\mathrm{cr}}=2$, and $2\pm\sqrt{2}$
($n_{\mathrm{GL}}=1$ in all cases);

- $t_{y}=\sqrt{2}$ (blue $\blacksquare$): $\mu_{\mathrm{cr}}=0$,
$2$ and $4$, the first has $n_{\mathrm{GL}}=2$ and the two latter
only show one gapless excitation in their energy dispersions;

- $t_{y}=2\sqrt{2}$ (green $\blacktriangle$): $\mu_{\mathrm{cr}}=2$
($n_{\mathrm{GL}}=2$) and $\mu_{\mathrm{cr}}=6$ ($n_{\mathrm{GL}}=1$).

Additionally, the quantum phase transitions for the systems with four
legs {[}Fig. \ref{fig:S_vs_mu_and_ty}(b){]} occur for the following
parameter values:

- $t_{y}=1$ (orange $\CIRCLE$): $\mu_{\mathrm{cr}}=\nicefrac{5\pm\sqrt{5}}{2}$
and $\mu_{\mathrm{cr}}=\nicefrac{3\pm\sqrt{5}}{2}$ (all the excitations
are gapless at a single point on each dispersion);

- $t_{y}=\nicefrac{4}{\sqrt{5}-1}$ (blue $\blacksquare$): $\mu_{\mathrm{cr}}=2$
($n_{\mathrm{GL}}=2$), and $\mu_{\mathrm{cr}}=6$, $(1\pm\sqrt{5})$
with $n_{\mathrm{GL}}=1$;

- $t_{y}=4$ (green $\blacktriangle$): $\mu_{\mathrm{cr}}=2\sqrt{5}$
(two gapless modes), and $\mu_{\mathrm{cr}}=(2\sqrt{5}\pm4)$ ($n_{\mathrm{GL}}=1$).

An interesting feature in the last case is that we can clearly see
that the entanglement entropy is sensitive to the number of gapless
modes of the energy bands. Notice that the middle peak at $\mu_{\mathrm{cr}}=2\sqrt{5}$
is higher than the one at a smaller chemical potential $\mu_{\mathrm{cr}}=2\sqrt{5}-4$.
We are going to examine this fact with more details in the next subsection
but before we move forward with this discussion let us look over the
behavior of $S$ as functions of $t_{y}$ for fixed $\mu$.

In Fig. \ref{fig:S_vs_mu_and_ty} we also display how the von Neumann
entropy behaves as the hopping mechanism along the rungs $t_{y}$
gets stronger. This is showed in the panels (c) and (d) for $N=3$
and $N=4$, respectively. This time we present calculations for several
different fixed values of chemical potential $\mu$. As we said before,
when there is no hopping strength along the rungs ($t_{y}=0$) the
legs of the ladder behave as $N$ independent Kitaev chains. Moreover,
in the absence of chemical potential each leg will contribute with
$N\,\ln2$ to entanglement entropy --- the horizontal dotted lines
in the graphs represent the values $S=N\,\ln2$. However, strong hopping
strength along the rungs $t_{y}$ weakens the quantum correlations
along the legs of the ladder as it increases the quantum correlations
between the left and right subsystems. Once again, we can notice jumps
in the von Neumann entropy profile and they are again signals of the
critical points corresponding to quantum phase transitions. The locations
of this divergences are in completely agreement with the phase diagrams
displayed in Fig. \ref{fig:critical-mu}.

For clarity, we will again list below the locations of the peaks observed
in the entanglement entropy displayed on panel \ref{fig:S_vs_mu_and_ty}-(c):

- $(\mu_{\mathrm{cr}},t_{y})=(0,\sqrt{2})$ in orange $\CIRCLE$ is
a critical point with $n_{\mathrm{GL}}=1$;

- $(\mu_{\mathrm{cr}},t_{y})=(1,\nicefrac{1}{\sqrt{2}})$, and $(1,\nicefrac{3}{\sqrt{2}})$
in blue $\blacksquare$, are both critical points that also present
single gapless mode dispersions;

- $(\mu_{\mathrm{cr}},t_{y})=(2,0)$, and $(2,2\sqrt{2})$ in green
$\blacktriangle$, the first corresponds to three disconnected chains
(three fully degenerated bands) and the latter has $n_{\mathrm{GL}}=2$.
It is worth to point that the entire line $\mu=\mu_{\mathrm{cr}}=2$
corresponds to a critical phase transition line, therefore, besides
at the two points mentioned here the line represents critical ladders
with $n_{\mathrm{GL}}=1$;

- $(\mu_{\mathrm{cr}},t_{y})=(4,\sqrt{2})$, and $(4,3\sqrt{2})$
in red $\blacktriangledown$, (single gapless mode excitations).

Last but not least, for the four-leg systems investigated in Fig.
\ref{fig:S_vs_mu_and_ty}(d) we found that at the peaks the dispersions
have the following features:

- $n_{\mathrm{GL}}=2$ for $(\mu_{\mathrm{cr}},t_{y})=(0,\nicefrac{4}{\sqrt{5}\pm1})$,
in orange $\CIRCLE$; 

- four fully degenerated bands (disconnected chains) at $(\mu_{\mathrm{cr}},t_{y})=(2,0)$,
and $n_{\mathrm{GL}}=1$ for $(\mu_{\mathrm{cr}},t_{y})=(2,\nicefrac{8}{\sqrt{5}\pm1})$
(blue $\blacksquare$);

- one gapless excitations at $(\mu_{\mathrm{cr}},t_{y})=(2\sqrt{5},6\pm2\sqrt{5})$,
and the peak at $(\mu_{\mathrm{cr}},t_{y})=(2\sqrt{5},4)$ corresponds
to the intersection of two critical lines (in green $\blacktriangle$).

This last curve also reveals that the entanglement entropy between
the subsystems is affected by the number of gapless modes in the energy
dispersion. The central peak at $(\mu_{\mathrm{cr}},t_{y})=(2\sqrt{5},4)$
with $n_{\mathrm{GL}}=2$ is higher than the other peaks that have
single gapless modes. The next subsection will be devoted to explore
how $S$ diverges at the quantum critical points. From several pasts
works we expect a logarithmic divergence.\cite{Vidal_Latorre,Vidal_Latorre2,CALABRESE_CARDY,CALABRESE_CARDY2,AFFLECK2,KOREPIN}
In fact we have also verified this behavior for the entanglement entropy
in the presence of interface defects.

\subsection*{Entanglement entropy with interface defects}

\begin{figure}
\begin{centering}
\includegraphics[width=0.5\textwidth]{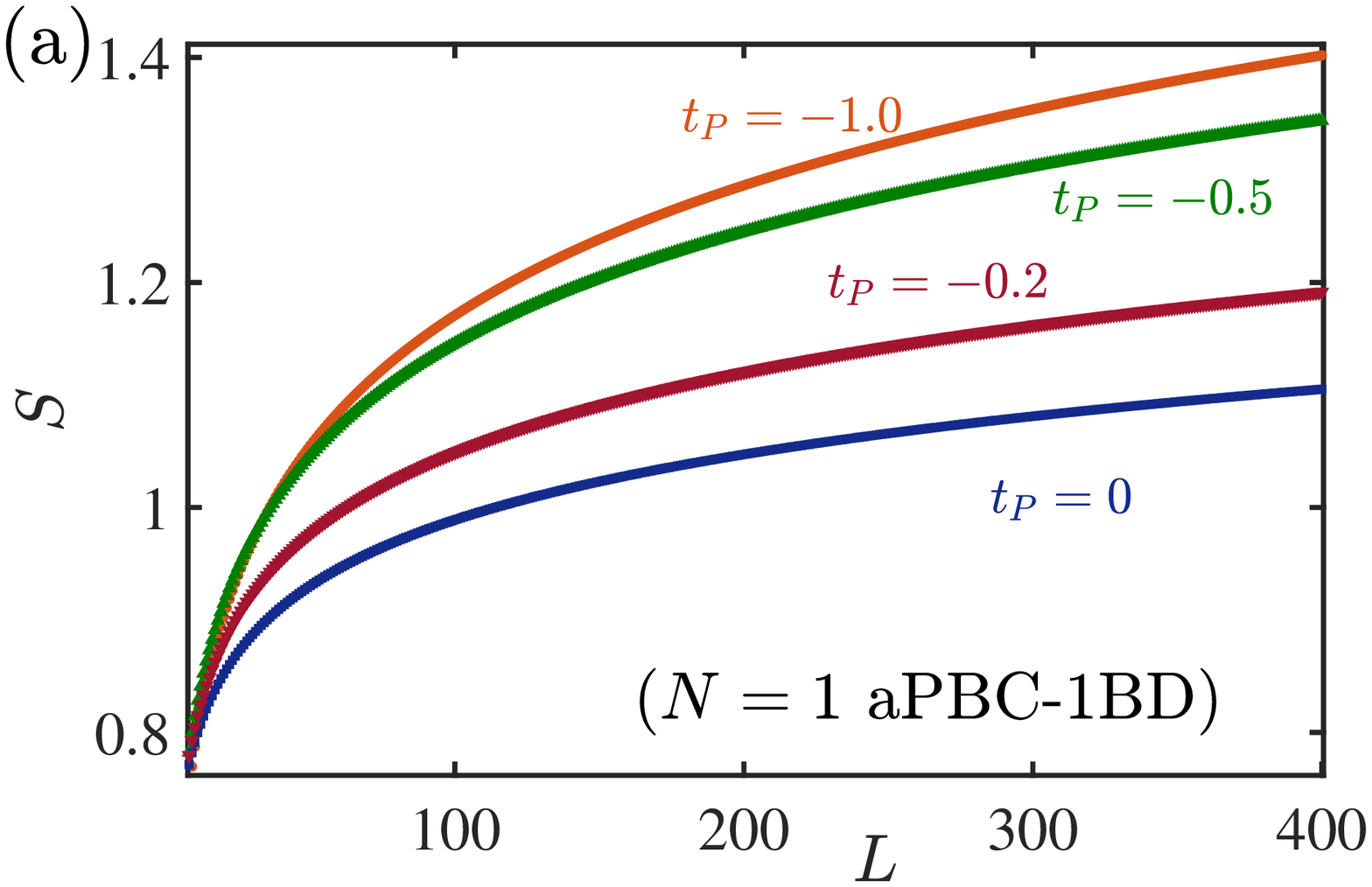}\includegraphics[width=0.5\textwidth]{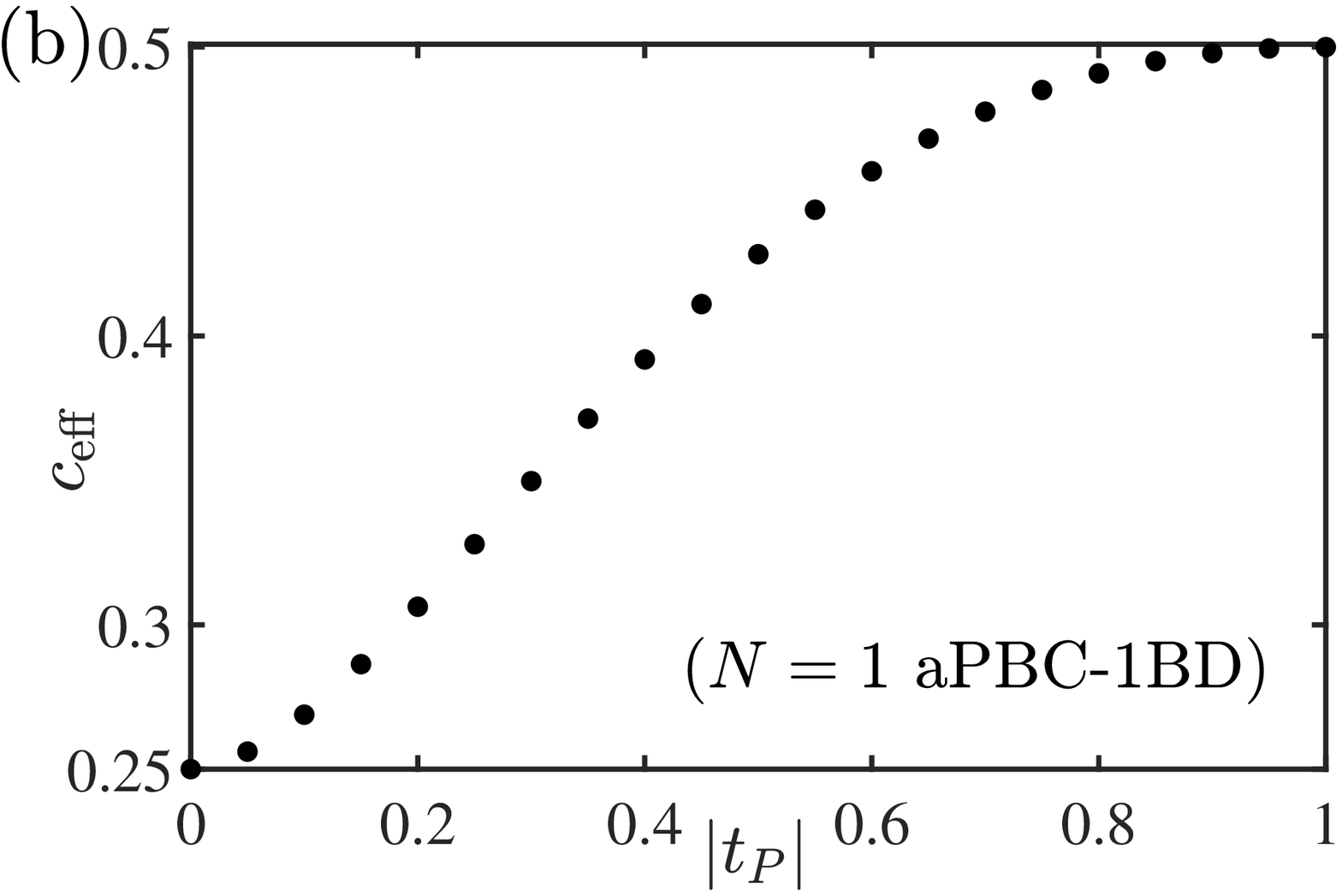}
\par\end{centering}
\caption{(a) Entanglement entropy $S$ versus the subsystem sizes $L$. We
fix $t_{0}=\gamma_{0}=1$, therefore we have one bound defect and
we set $\mu=|\mu_{\mathrm{cr}}|=2$. We also set $t_{P}=\gamma_{P}$
and show $S$ for different values of $t_{P}.$ (b) Effective central
charge as a function of the defect strength. \label{fig:Kitaev N=00003D1}}
\end{figure}
In this section we will go beyond the homogeneous ladders discussed
in the last section. Here we have not only analyzed the universal
logarithmic divergence of $S$ for homogeneous ladders but we have
also considered the ground state entanglement entropy between the
two subsystems in the \textit{presence of interface defects}. We have
considered the cases of bond defects at the boundaries of the sub-ladder
--- this was done by changing ($t_{0},\gamma_{0}$) and/or ($t_{P},\gamma_{P}$).
We found a logarithmic behavior for the von Neumann entanglement entropy
with constants depending continuously on the defect strength.

In order to fit our data we have used the following expression
\begin{equation}
S=\eta\frac{c_{\mathrm{eff}}}{6}\ln L+\beta+\frac{\alpha}{L},\label{eq:S=00003DlogL}
\end{equation}
where the prefactor $c_{\mathrm{eff}}$ of the logarithmic leading
term plays the role of the ECC. $\eta$ is the number of connecting
points between the subsystems, i.e., if the homogeneous ladder has
OBC along the chains we set $\eta=1$ and if PBC or anti-PBC are imposed
we choose $\eta=2$. $\beta$ and $\alpha$ are two non-universal
constants, moreover the latter is used to take into account sub-leading
corrections of the entanglement entropy. It is important to stress
that the logarithm leading finite-size correction of the behavior
of $S$ is expected to be asymptotic, i.e., it does not hold for small
$L$. In a practical way we have neglected the data for $L\lesssim100/N$
when we used Eq. (\ref{eq:S=00003DlogL}) to obtain $c_{\mathrm{eff}}$.

In the following subsection we are going to examine separately the
Kitaev ladders with a non zero superconducting pairing potential ($\gamma\neq0$)
as well as the case with $\gamma=0$. 

\subsubsection*{$\gamma\protect\neq0$ --- $N=1$: single chain}

Let us start our investigation with a single Kitaev critical chain.
Igl{\'o}i \textit{et al.} \cite{Igl_i_2007} did a similar discussion
of this simple one-dimensional case in 2007. It is a very well known
fact, and we have discussed that in Section \ref{sec:Gapless-points},
that the critical chemical potential is $|\mu_{\mathrm{cr}}|=2$,
so we set $\mu=2$ here. As we have said before, the homogeneous chain
is analogous to the famous Ising chain in a critical transverse magnetic
field, therefore the central charge is exactly $c=1/2$. Furthermore,
we are going to impose anti-PBC to the chain, so there are two points
connecting the subsystems and we have to set $\eta=2$ in Eq. (\ref{eq:S=00003DlogL}).
Moreover, we have considered the influence of one bound defect (1BD),
we do it by setting $t_{0}=1=\gamma_{0}$ and we calculate the entanglement
entropy changing the defect strengths, $t_{P}$ and $\gamma_{P}$,
at the other boundary, we also fix $t_{P}=\gamma_{P}$.

In Fig. \ref{fig:Kitaev N=00003D1}(a) we show the behavior of the
von Neumann entanglement entropy $S$ as a function of the subsystems
size $L$ for four different values of $t_{P}$. As expected, $S$
is a increasing function of $L$, in fact $S$ diverges with a logarithmic
behavior --- the system is at a quantum phase transition. For $t_{P}=-1$,
the homogeneous chain case, we get $c_{\mathrm{eff}}\approx0.5000019$.
Therefore, within our numerical precision (finite $L$) we can say
that $c_{\mathrm{eff}}=c=1/2$. As we decrease the defect strength
the effective central charge also decreases. We found $c_{\mathrm{eff}}\approx0.4283$
for $t_{P}=-0.5$, $c_{\mathrm{eff}}\approx0,3063$ for $t_{P}=-0.2$,
and $c_{\mathrm{eff}}\approx0.25002$ for $t_{P}=0$. So, we have
found here that the ECC for $t_{P}=0$ is (approximately) half of
the central charge $t_{P}=-1$. This fact is not a surprise because
$t_{P}=0$ corresponds to a homogeneous chain with OBC, so its central
charge is actually $c=1/2$ when we use Eq. (\ref{eq:S=00003DlogL})
with $\eta=1$ connecting points between the two sub-chains.

Additionally, we have calculated $S$ for other values of $t_{P}$.
In the panel (b) of Fig. \ref{fig:Kitaev N=00003D1} we show how the
effective central $c_{\mathrm{eff}}$ decreases as the hopping term
and superconducting gap, at one of the subsystems boundary ($|t_{P}|=|\gamma_{P}|$),
decrease.

\subsubsection*{$\gamma\protect\neq0$ --- $N>1$: multi-leg Kitaev ladders}

\begin{figure}
\begin{centering}
\includegraphics[width=0.5\textwidth]{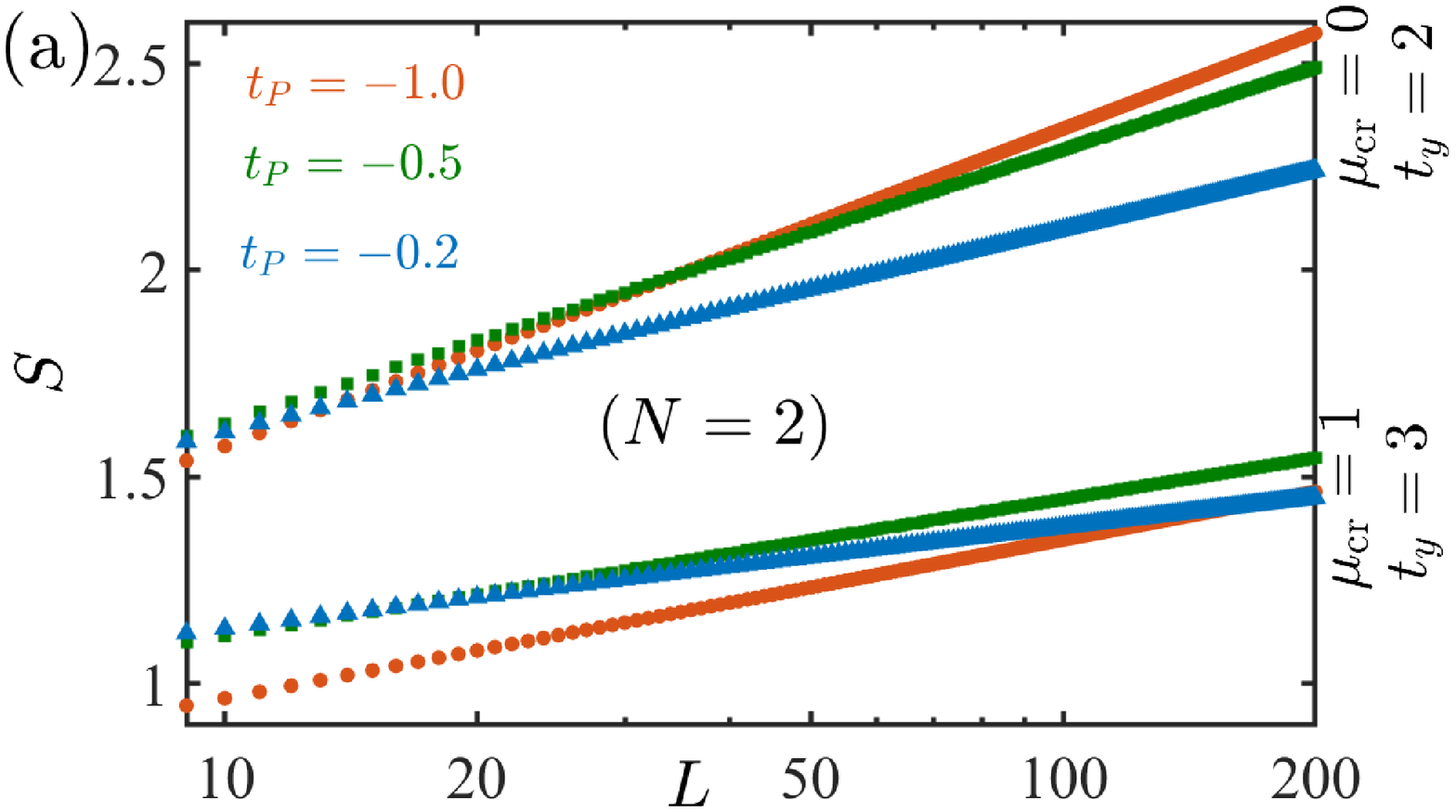}\includegraphics[width=0.5\textwidth]{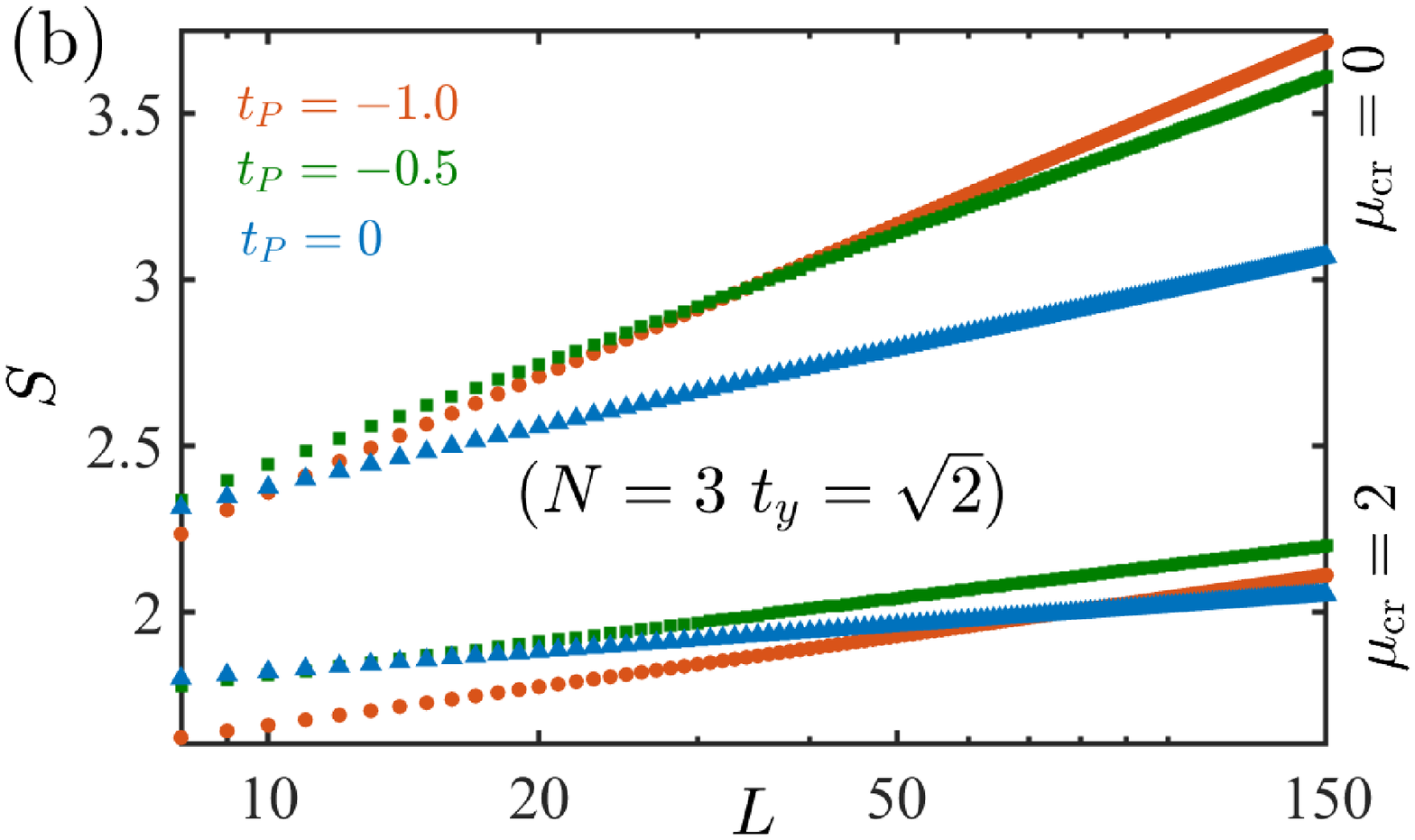}
\par\end{centering}
\begin{centering}
\includegraphics[width=0.5\textwidth]{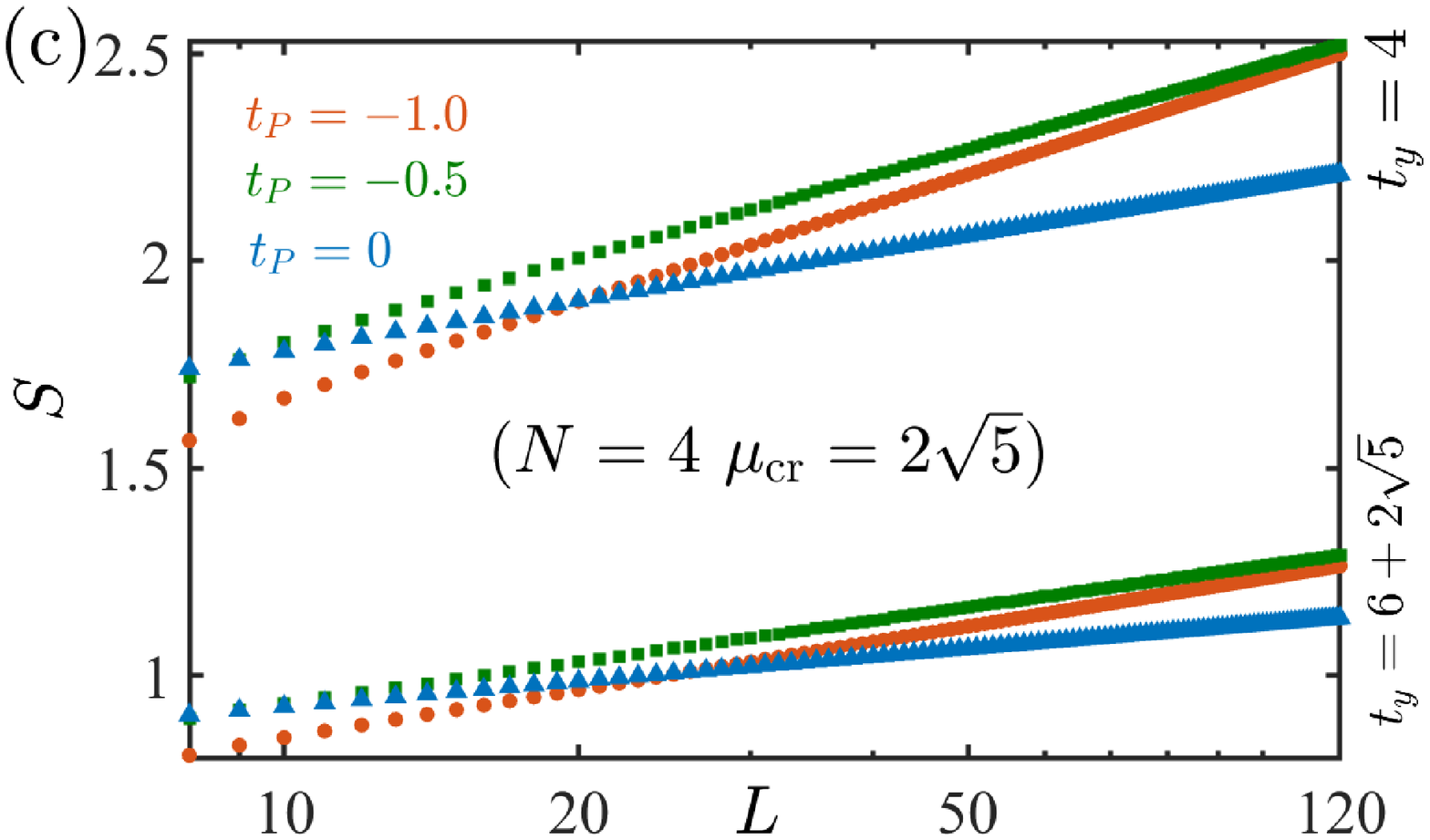}
\par\end{centering}
\caption{Bipartite von Neumann entanglement entropy $S$ as a function of the
composing ladders size $L$ (log scale). We have fixed $t_{0}=\gamma_{0}=1$
and set $t_{P}=\gamma_{P}$ here. The number of legs are $N=2$, $N=3$,
and $N=4$ in the panels (a), (b), and (c), respectively. Finally,
these graphs helped us to conclude that: since all the ladder considered
are at critical points, $S$ diverges as $S\propto c_{\mathrm{eff}}\ln L$,
moreover $c_{\mathrm{eff}}$ is proportional to the number of gapless
modes in the energy dispersion. \label{fig:Kitaev N>1}}
\end{figure}
The discussion of the cases $N>1$ can be divided into \textit{disconnected
chains }and \textit{connected chain}. The first one is the trivial
case where we set the hopping terms along rungs to zero ($t_{y}=0$).
In this situation, the critical chemical potential is $|\mu_{\mathrm{cr}}|=2$.
Moreover, since we have $N$ disconnected chains we expect that the
entanglement entropy $S$ is just $N$ times the entanglement entropy
of  a single chain. In fact, that is the result that we have obtained
and consequently the ECC we got is also $N$ times the $c_{\mathrm{eff}}$
obtained for $N=1$. 

The second situation is $t_{y}\neq0$, so the chains are connected
through inter-leg hopping, this is a more interesting case and we
will devote the remaining of this subsection to discuss this. Once
more, we have imposed anti-PBC along the chains that compose the ladders
and we have investigated systems with bound defects at one of the
boundaries of the subsystems. This was implemented by again fixing
$t_{0}=1=\gamma_{0}$ and changing $t_{P}$, but keeping $t_{P}=\gamma_{P}$,
at the other boundary. 

In Fig. \ref{fig:Kitaev N>1}(a) we present the behavior of $S$ as
function of $L$ for six critical two-leg ladders. The three lower
curves show the entanglement entropy of ladders with $\mu_{\mathrm{cr}}=1$
and $t_{y}=3$ and three different values of $t_{P}$. As we have
previously discussed {[}see for example Fig. \ref{fig:critical-mu}{]}
these set of parameters lead to dispersions with one gapless mode
excitations. We have fitted these data to Eq. (\ref{eq:S=00003DlogL})
in order to estimate the ECC. We have obtained $c_{\mathrm{eff}}\approx0.5000077$
for $t_{P}=-1$, $c_{\mathrm{eff}}\approx0.4284$ for $t_{P}=-0.5$,
and $c_{\mathrm{eff}}\approx0.3065$ for $t_{P}=-0.2$. We have also
evaluated the ECC for other $t_{P}$ values and we found that they
are equal (within our numerical precision) to the values of $c_{\mathrm{eff}}$
we obtained for $N=1$. Additionally, the three upper curves of Fig.
\ref{fig:Kitaev N>1}(a) correspond to ladders with $\mu_{\mathrm{cr}}=0$
and $t_{y}=2$ and three $t_{P}$ values. According to the phase diagram
presented in Fig. \ref{fig:critical-mu}, for this values of $\mu_{\mathrm{cr}}$
and $t_{y}$ the energy dispersion of the critical ladder has two
gapless mode excitations at both $k=0$ and $|k|=\pi$. When we fitted
these data to Eq. (\ref{eq:S=00003DlogL}) to evaluate $c_{\mathrm{eff}}$
we found that $c_{\mathrm{eff}}\approx1.000015$ for $t_{P}=-1$,
$c_{\mathrm{eff}}\approx0.8567$ for $t_{P}=-0.5$, and $c_{\mathrm{eff}}\approx0.6130$
for $t_{P}=-0.2$. We have once again contrasted these values to the
ones we got for the single chain, and taking into account the numerical
precision, the comparisons conduct us to concluded that the ECC here
is double the one found for $N=1$.

For the sake of completeness we also illustrate how the bipartite
von Neumann entanglement entropy behaves for ladders composed by $N=3$
and $N=4$ legs in Fig. \ref{fig:Kitaev N>1}(b) and (c), respectively.
For $N=3$, we have depicted data for $t_{y}=\sqrt{2}$ and two different
critical chemical potentials $\mu_{\mathrm{cr}}=2$ ($n_{\mathrm{GL}}=1$)
and $\mu_{\mathrm{cr}}=0$ ($n_{\mathrm{GL}}=2$). The ECC, as a function
of $t_{P}=\gamma_{P}$, that we have found is the same we found for
$N=2$. In other words, $c_{\mathrm{eff}}$ is the same showed in
Fig. \ref{fig:Kitaev N=00003D1}(b) times the number of gapless modes
presented in the energy dispersion. The same general results were
perceived for the ladders with four chains. The parameters in Fig.
\ref{fig:Kitaev N>1}(c) are $(\mu_{\mathrm{cr}},t_{y})=(2\,\sqrt{5},6+2\,\sqrt{5})$
and $(\mu_{\mathrm{cr}},t_{y})=(2\,\sqrt{5},4)$, and these systems
have $n_{\mathrm{GL}}=1$ and $n_{\mathrm{GL}}=2$, respectively. 

It is import to comment here that several other critical ladders were
considered, and, the above general results were always verified within
our numerical precision. Therefore, we conclude that for connected
Kitaev chains with $\gamma\neq0$ (and in a rectangular geometry)
we should expect that the conclusions we found here should always
be true. It is worthy to point that we have only showed the behavior
of $c_{\mathrm{eff}}(t_{P})$ in the range $0\leqslant|t_{P}|\leqslant1$
because we have noticed that the system has the following symmetries
$c_{\mathrm{eff}}(t_{P})=c_{\mathrm{eff}}(-t_{P})=c_{\mathrm{eff}}(t_{P}^{-1})$. In the next section we are going to consider the case without
the superconducting paring amplitude. 

\subsubsection*{$\gamma=0$ : multi-leg quadratic hamiltonian}

\begin{figure}
\begin{centering}
\includegraphics[width=0.5\textwidth]{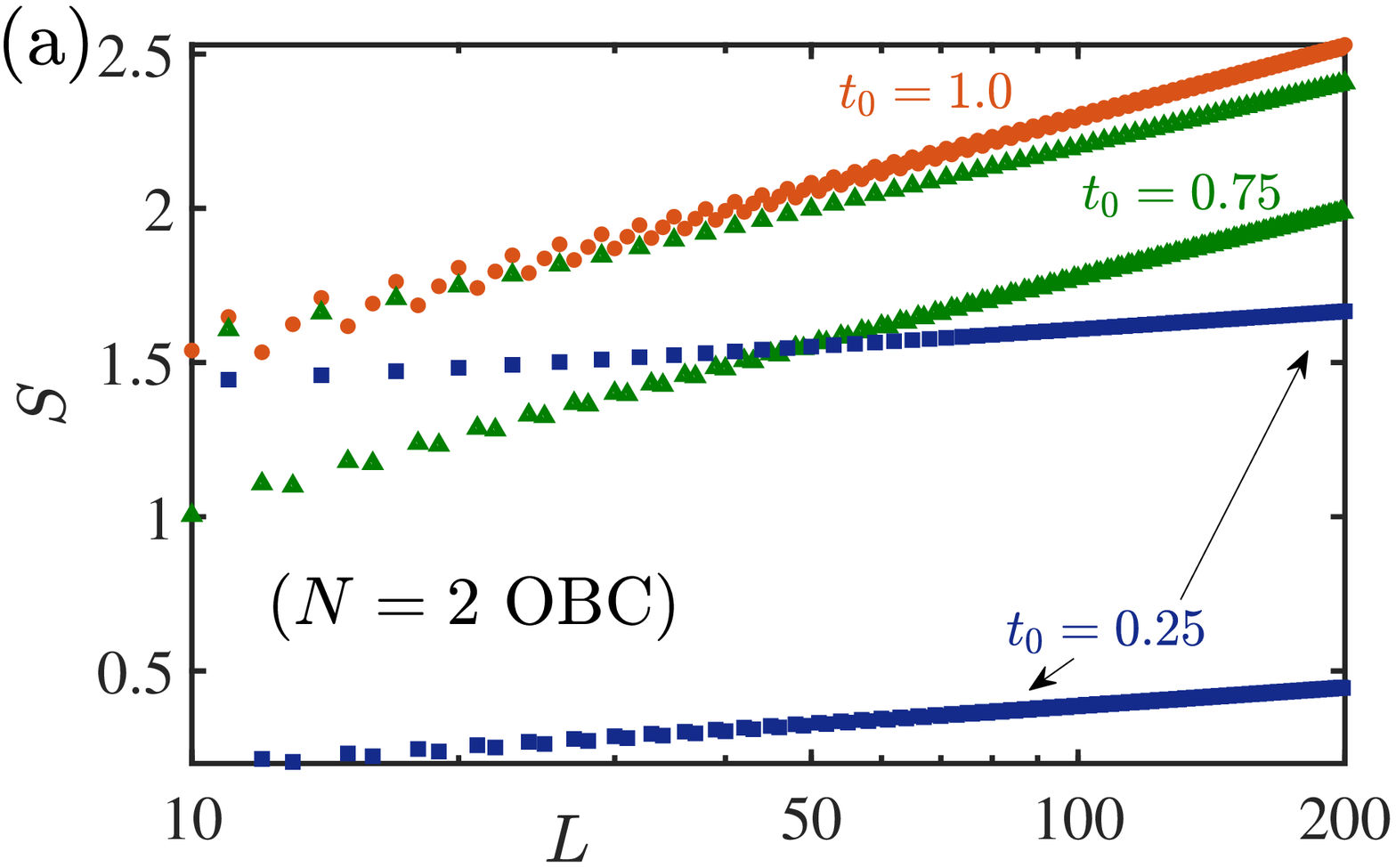}\includegraphics[width=0.5\textwidth]{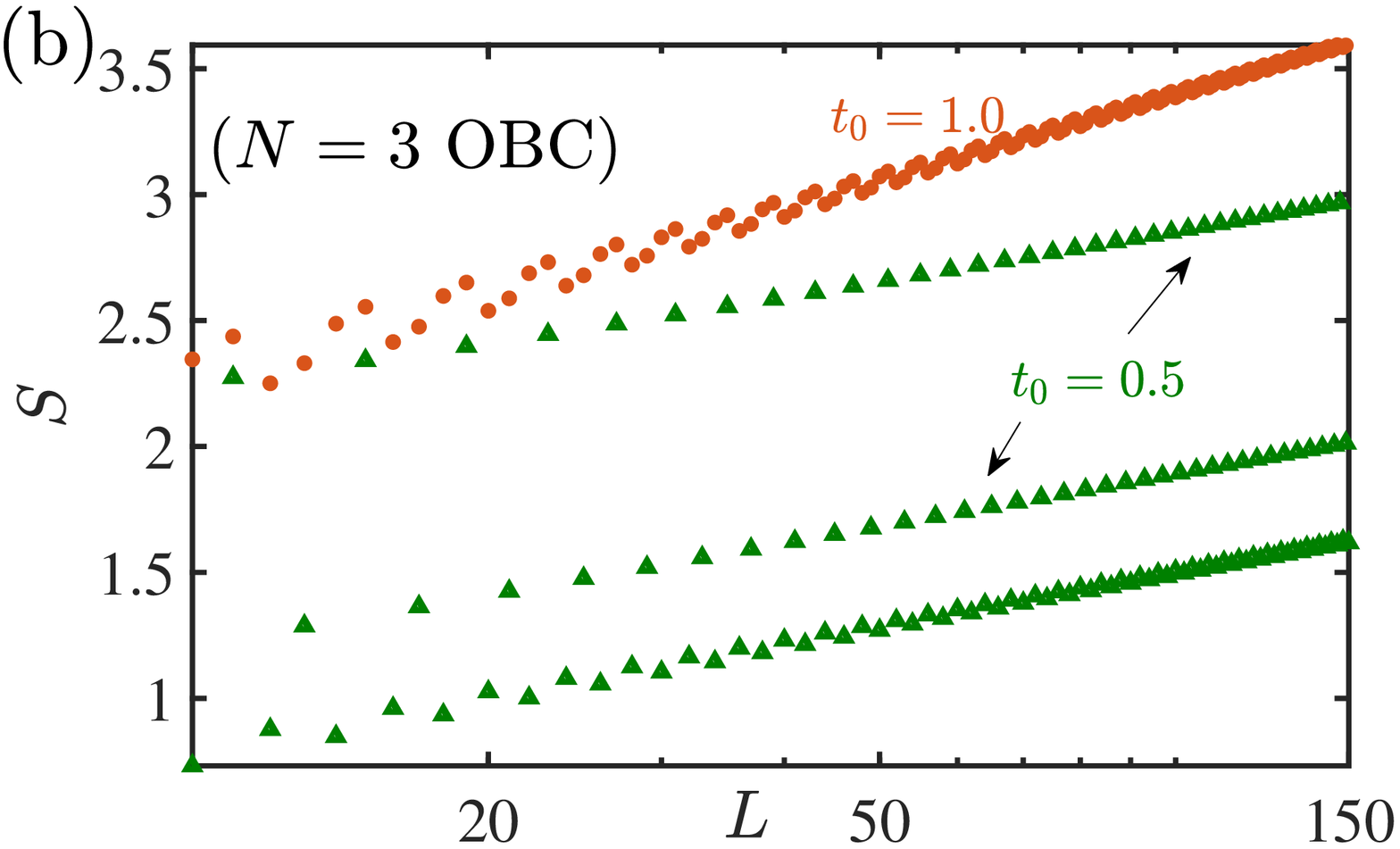}
\par\end{centering}
\begin{centering}
\includegraphics[width=0.5\textwidth]{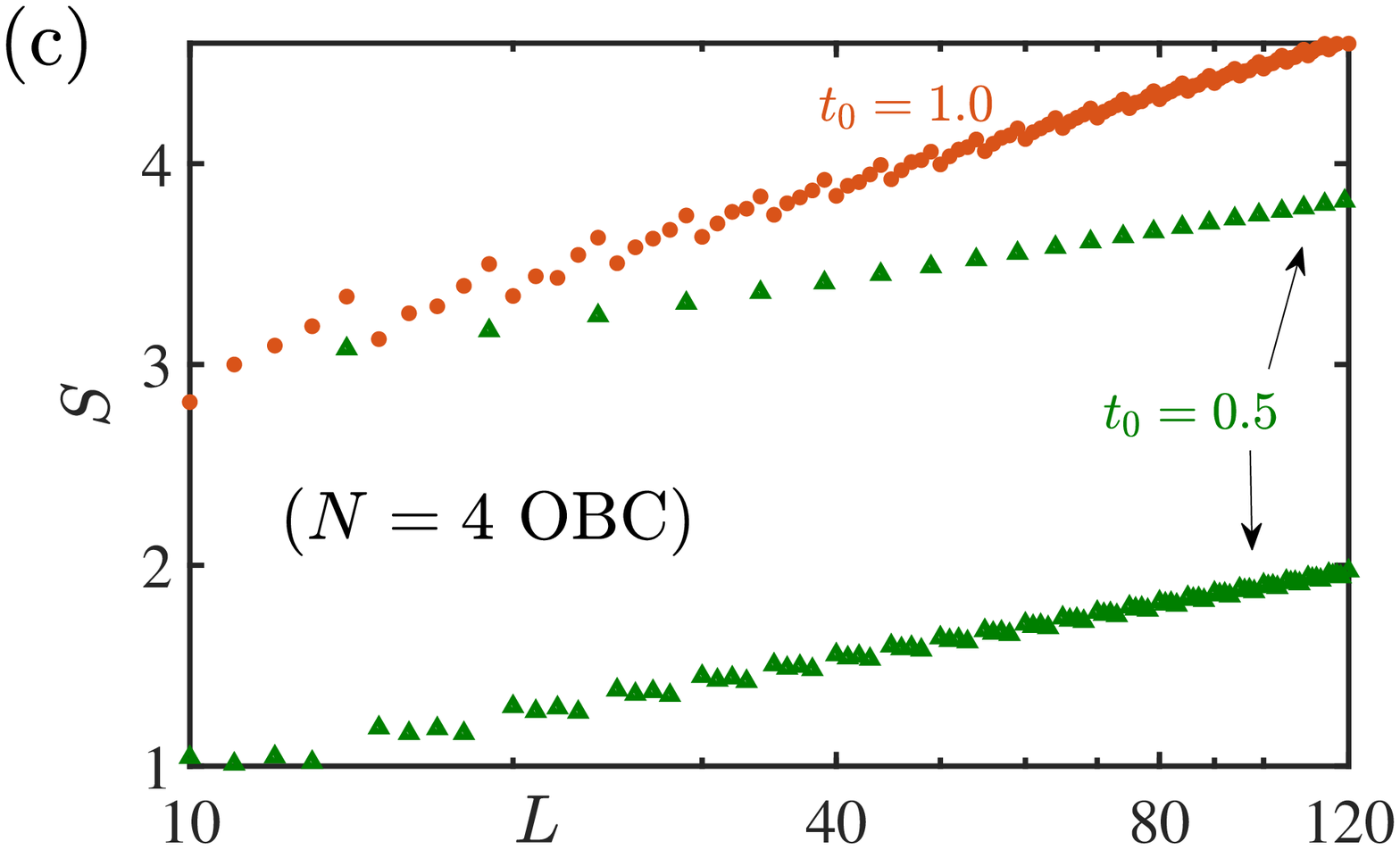}\includegraphics[width=0.5\textwidth]{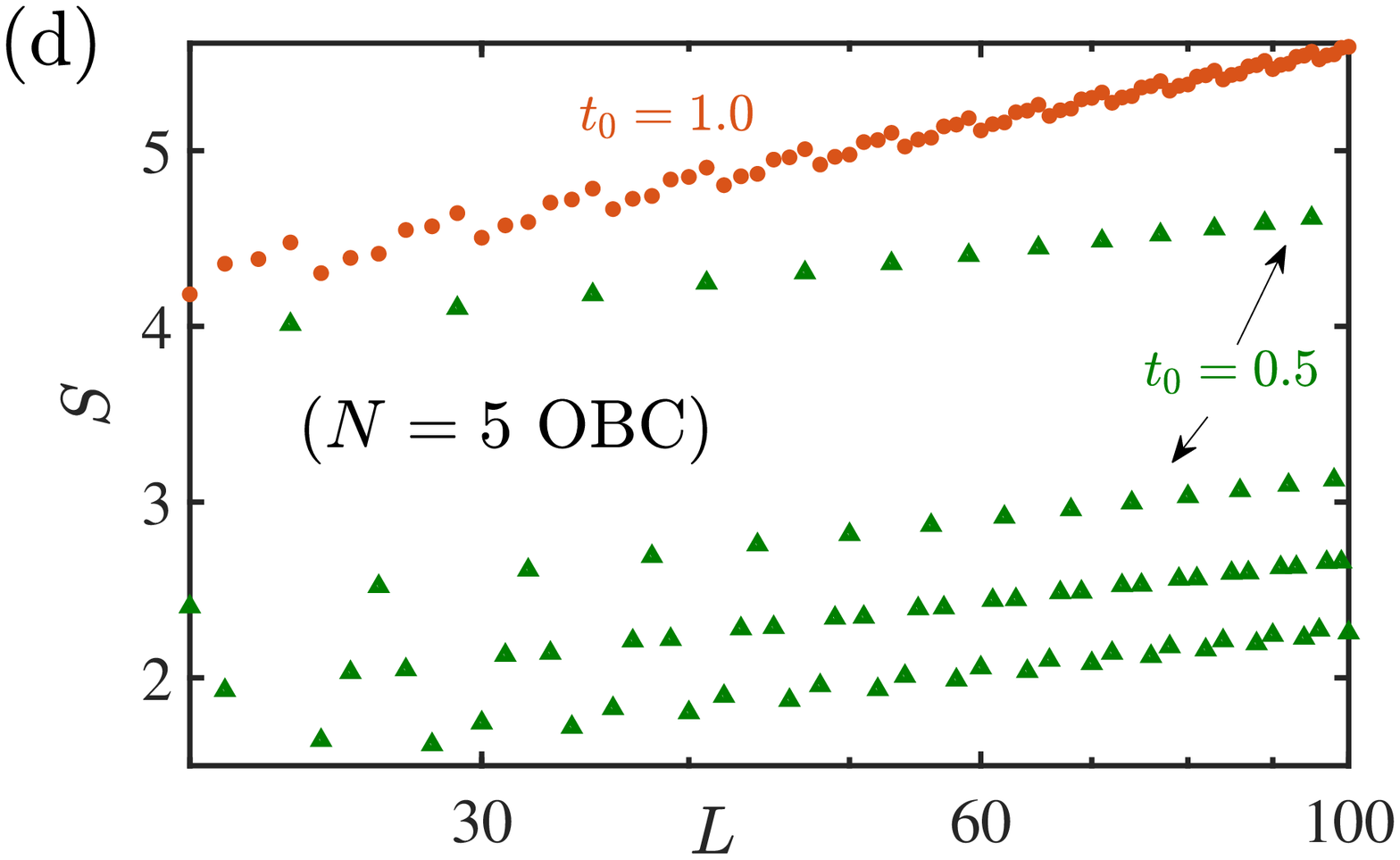}
\par\end{centering}
\caption{(a) - (d) Entanglement entropy for quadratic hamiltonians ($\gamma_{n}=0$)
with OBC as a function of the size $L$ (log scale) of the subsystems.
We show results for $N=2,3,4$ and $5$ and different defect hopping
amplitude $t_{0}$ between the sub-ladders. \label{fig:S_gamma=00003D0-OBC}}
\end{figure}
Our main goal here is to investigate \textit{non-homogeneous} situations
across the interfaces in the absence of superconducting gap. The homogeneous
cases were already discussed in the past,\textcolor{red}{\cite{Xavier_2014}}
and it is known that the ECC is also proportional to $n_{\mathrm{GL}}$.
But, different from what we have observed in the previous subsection
--- where we found that for $\gamma\neq0$ we can only tune critical
ladders of connected legs with one or two gapless modes --- for $\gamma=0$
and more than two connected legs ($N>2$) it is possible to obtain
ladders with $n_{\mathrm{GL}}>2$. In this subsection, we have set
the hopping along the rungs as $t_{y}=1=t_{x}$ and we calculate $S$
for several values of $t_{0}$ and $t_{P}$.

Let us start our discussion of inhomogeneities reviewing the case
of a single leg/chain ($N=1$) with OBC (so we fix $t_{P}=0$). Eisler
\textit{et al.}\textcolor{red}{{} \cite{Eisler_2015}} found that the
behavior of the bipartite entropy versus the size $L$ of the subsystems
has oscillations between odd and even sizes. They were able to estimate
the ECC using almost the same procedure employed before for finite
$\gamma$. The difference is that they needed to fit the odd and even
$L$ values {[}using the Eq. (\ref{eq:S=00003DlogL}) with $\eta=1${]}
separately. They found that $c_{\mathrm{eff}}$ continuously increases
from zero (two sub-chains are disconnected) to $c_{\mathrm{eff}}\approx1$
when $t_{0}$ increases from zero to one, as well.\footnote{For $t_{0}=1$, we have the homogeneous OBC system and $c_{\mathrm{eff}}$
is expected to be exactly equal one, in the thermodynamic limit. Moreover
this situation corresponds to the famous critical XX spin-$1/2$ chain.}

Results for ladder with $N=2,3,4$ and $5$ are presented in Figs.
\ref{fig:S_gamma=00003D0-OBC}(a) to (d), where we show the behavior
of the bipartite entropy versus the size $L$ of the sub-ladder for
different values of the defect hopping amplitude $t_{0}$. Once more,
there are some oscillatory pattern on the behavior of the entanglement
entropy. For example, for $N=2$, panel (a), we noticed that $S(L)$
splits into three ($2+1$) different curves, moreover, all sizes that
are multiple of three {[}$L\equiv0$ mod($3$){]} belong to the same
branch, and also all the sizes that are multiple of three plus one
{[}$L\equiv1$ mod($3$){]} belong to another branch, and finally
the sizes that are multiple of three plus two {[}$L\equiv2$ mod($3$){]}
belong to the last branch.

\begin{figure}
\begin{centering}
\includegraphics[width=0.5\textwidth]{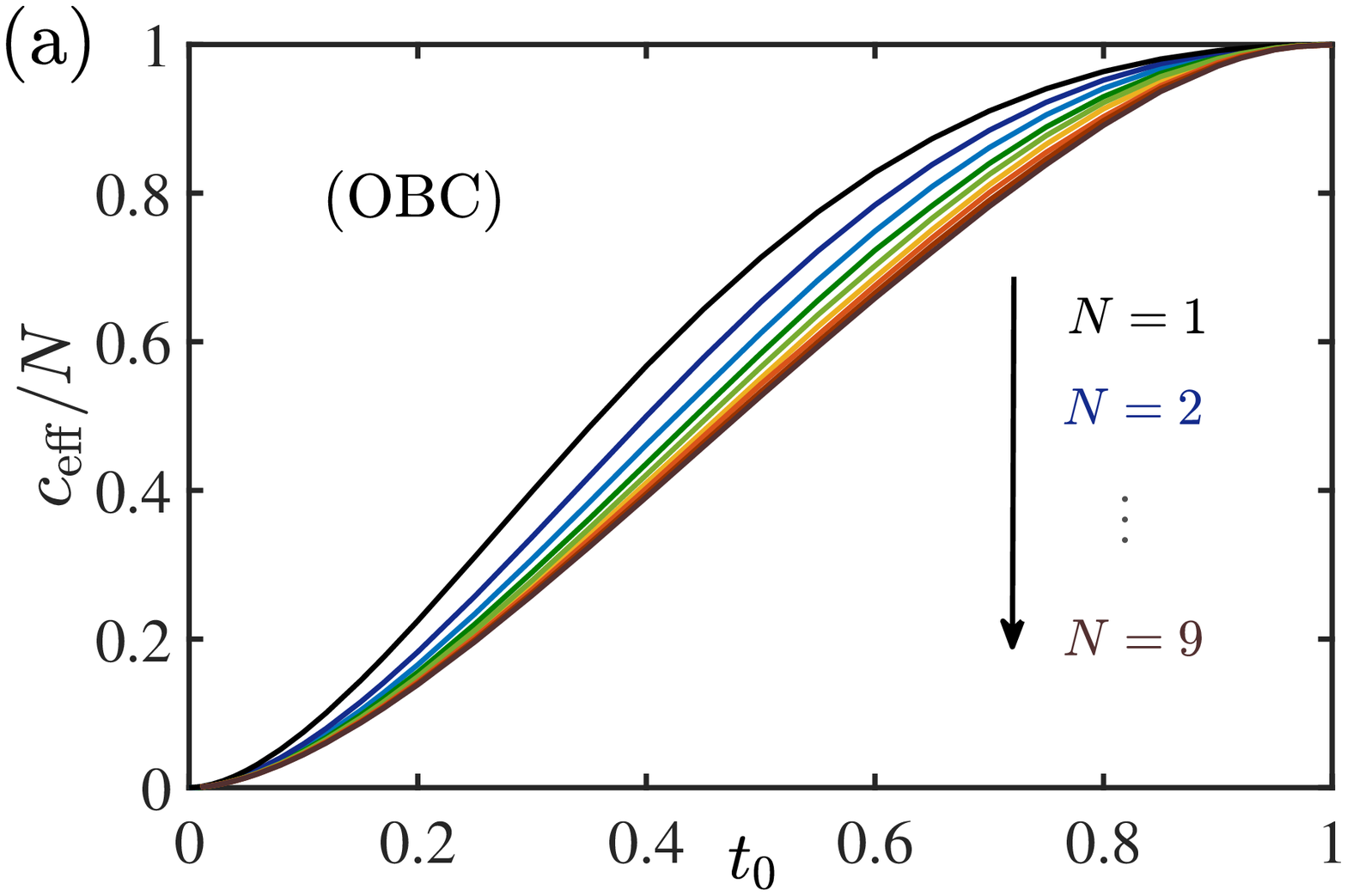}\includegraphics[width=0.5\textwidth]{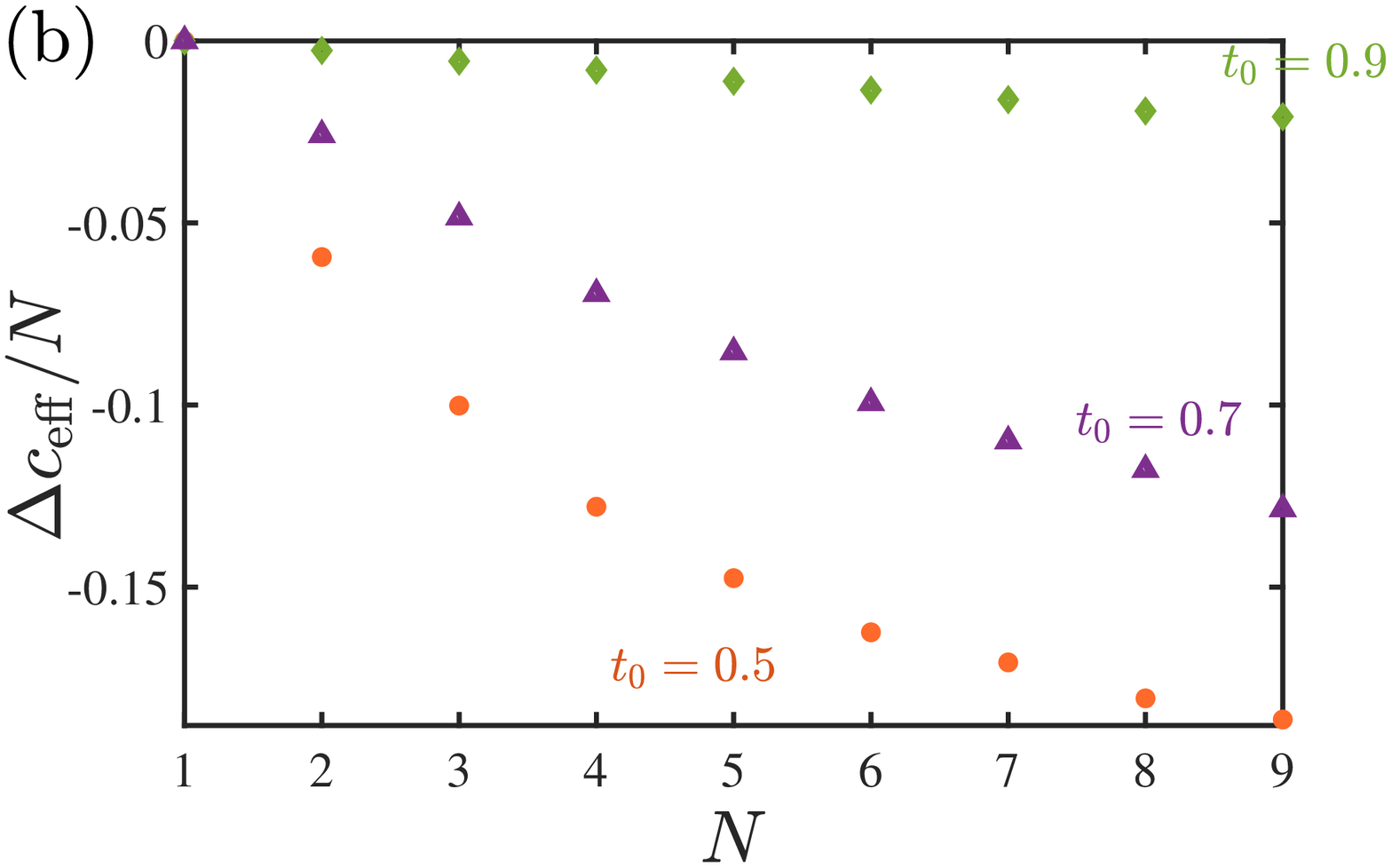}
\par\end{centering}
\caption{(a) The behavior of the effective central charge versus $t_{0}$ for
$N\leqslant9$. We have set $t_{y}=t_{x}=1$. (b) The effective central
charge shortage $\Delta c_{\mathrm{eff}}/N=\ensuremath{c_{\mathrm{eff}}(N)/N-c_{\mathrm{eff}}(1)}$
as a function of the number of legs for the cases investigated in
Fig. \ref{fig:S_gamma=00003D0-OBC}.\label{fig:ceff_vs_N}}
\end{figure}

For the three-leg ladders {[}Fig. \ref{fig:S_gamma=00003D0-OBC}(b){]},
we can observe that $S(L)$ for $L=1,5,9,13,\cdots$ follow the same
logarithmic behavior, and that the sets $\{L=2,6,10,14,\cdots\}$,
$\{L=3,7,11,15,\cdots\}$, and $\{L=4,8,12,16,\cdots\}$ belong to
different branches. In general we found that for a given number of
legs $N$ there will be ($N+1$) different branches on the behavior
of $S(L)$. Moreover, each branch will have a value of $c_{\mathrm{eff}}$.
In general, the average among these values was taken in order to estimate
the ECC and we show these results in Fig. \ref{fig:ceff_vs_N}(a).
We show data for the cases of $N=1$ to $N=9$. For the sake of explicitness,
in order to calculate the ECC for $N=1,$$2$, $3$, $4$, $5$, $6$,
$7$, $8$, and $9$, the maximum sizes we have considered were $\mathrm{max(}$$L)=400,$
$200$, $150$, $120$, $100$, $90$, $80$, $75$, and $70$, respectively.
The top (black) line in this panel corresponds to the $c_{\mathrm{eff}}$
of a chain and it is in agreement with results found on Ref.\textcolor{red}{{}
\cite{Eisler_2015}}. For a general $N$, we observe that the ECC
also continuously increases from $c_{\mathrm{eff}}=0$ but reaches
$c_{\mathrm{eff}}=N$ as we increase $t_{0}$ in the range $[0,1]$
--- notice that, $t_{0}=1$ corresponds to the homogeneous OBC system,
therefore, $c_{\mathrm{eff}}$ is expected to be equal to the number
of gapless modes in the dispersion $\Lambda_{k,q}=\pm\epsilon_{k}$,
and for the parameters that we have used this is equal to the number
of legs of the ladder ($n_{\mathrm{GL}}=N$). 

It is worthy to point out here that, for the case $\gamma\neq0$,
the ECC for systems with two gapless modes was exactly equal twice
the ECC of a single chain (one gapless mode) for $0\leqslant t_{0}\leqslant1$.
On the other hand, this does not happen in the absence of the superconducting
paring potential. Moreover, for $\gamma=0$, our data show $c_{\mathrm{eff}}/N$
decreases as the number of legs $N$ increases for $t_{0}\neq0$ and
$t_{0}\neq1$. In other words, in the presence of a defect hopping
amplitude $0<t_{0}<1$ the ECC of $N$ connected legs ($t_{y}=1$)
is smaller than the $c_{\mathrm{eff}}$ for $N$ disconnected chains
($t_{y}=0$), because for the latter $c_{\mathrm{eff}}/N$ does not
depend on $N$. Therefore, we expect that as the hopping mechanism
along the rungs increases the ECC of the bipartite entropy between
the left and right rungs decreases. Putting it another way, $t_{y}\neq0$
should decrease the quantum correlation along legs. We show in Fig.
\ref{fig:ceff_vs_N}(b) this shortage {[}$\Delta c_{\mathrm{eff}}=\ensuremath{c_{\mathrm{eff}}(N)-Nc_{\mathrm{eff}}(1)}${]}
of the effective central charge as a function of the number of legs
for some values of $t_{0}$. Moreover, we have observed that $\Delta c_{\mathrm{eff}}$
is minimum, for a given $N$, when the defect strength is $t_{0}=0,5$.

\begin{figure}
\begin{centering}
\includegraphics[width=0.5\textwidth]{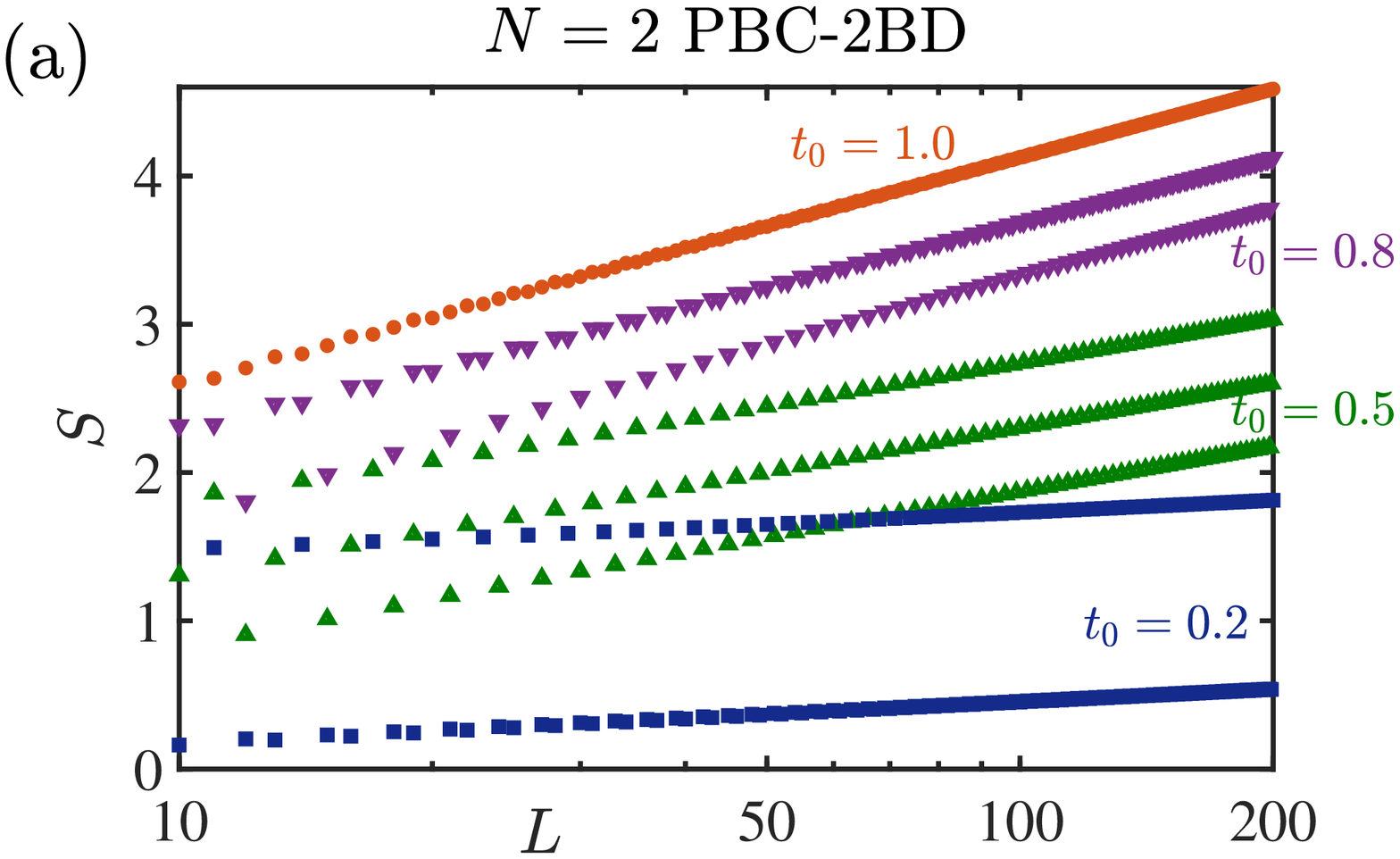}\includegraphics[width=0.5\textwidth]{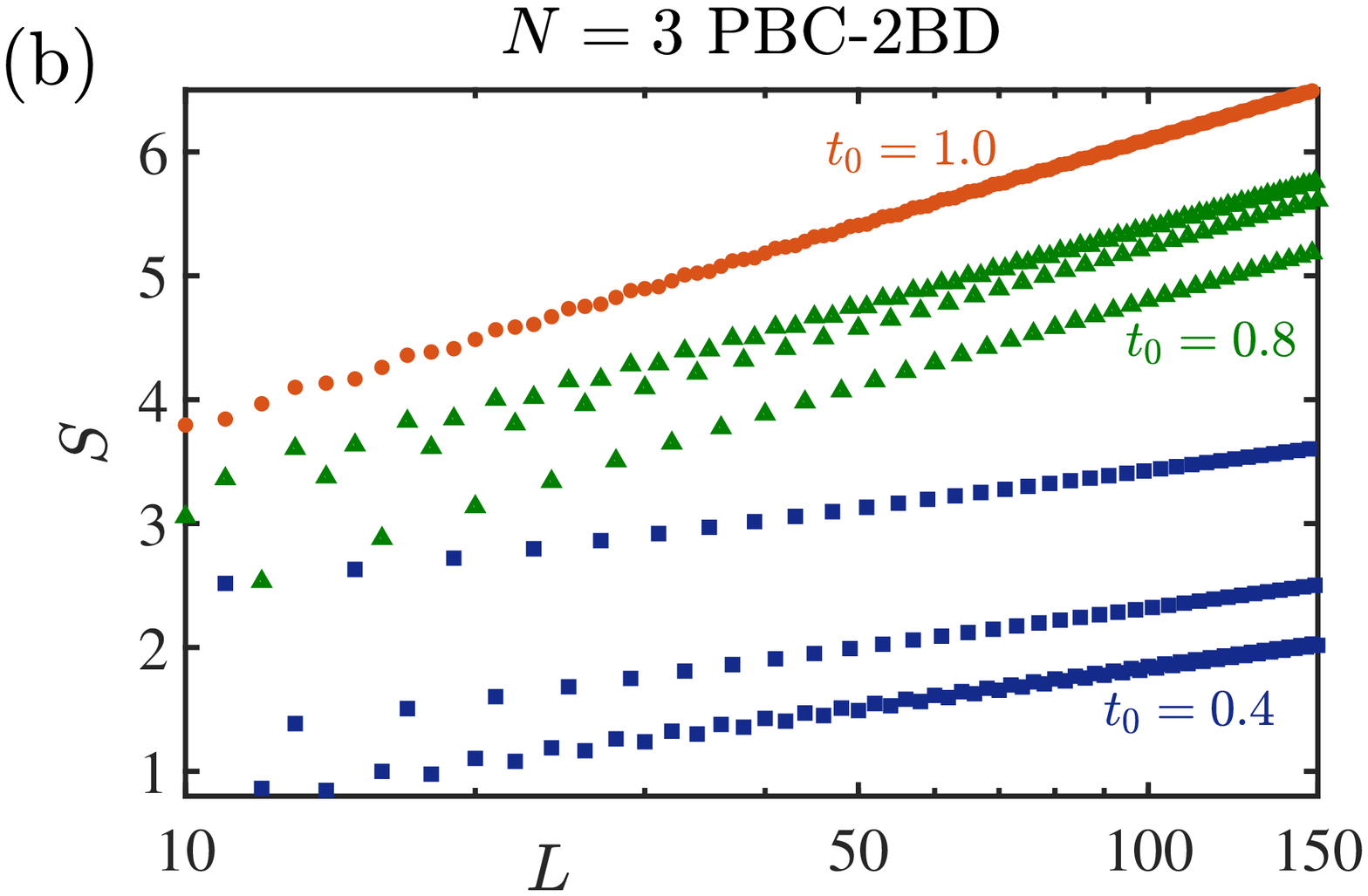}
\par\end{centering}
\begin{centering}
\includegraphics[width=0.5\textwidth]{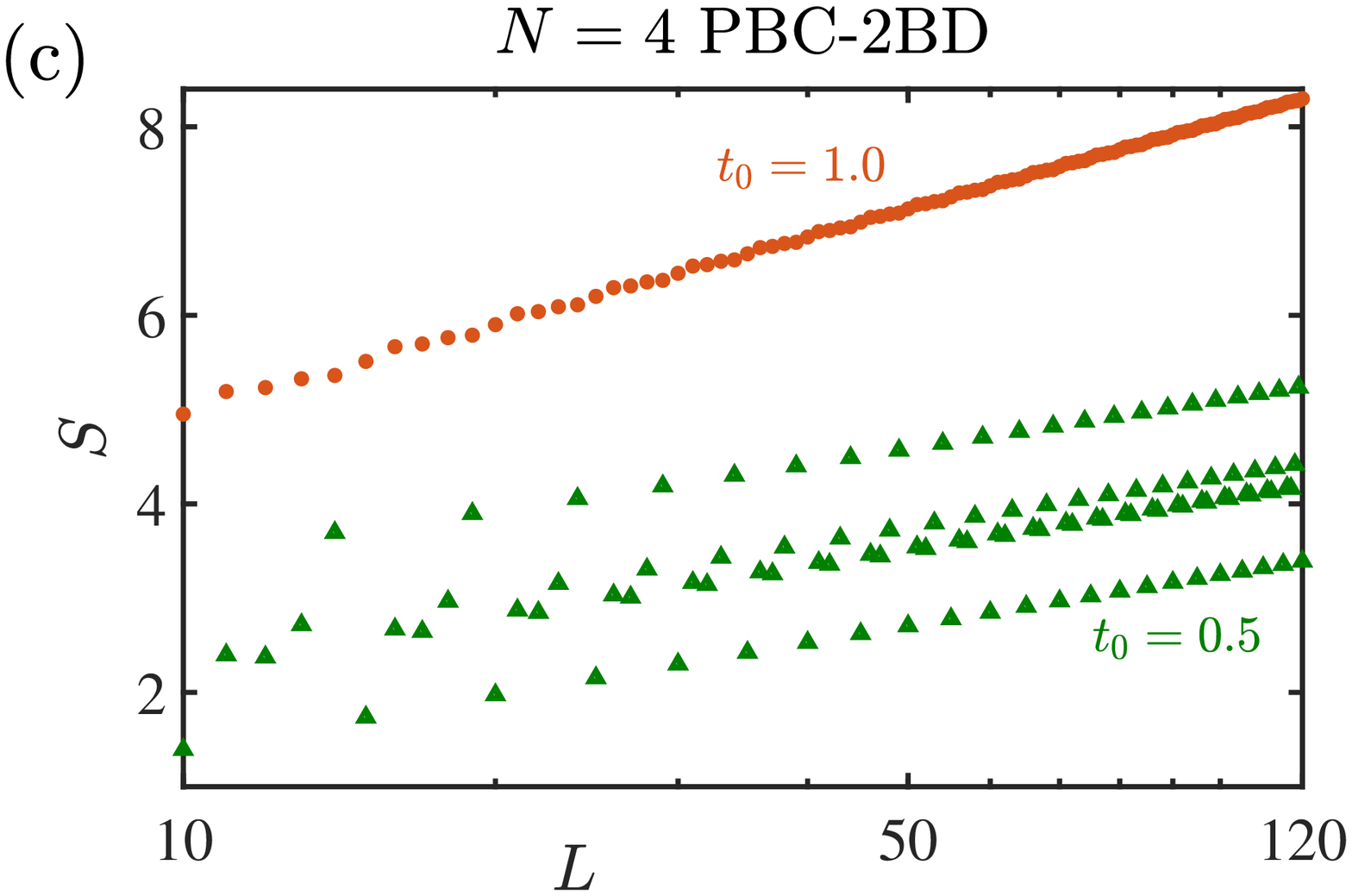}\includegraphics[width=0.5\textwidth]{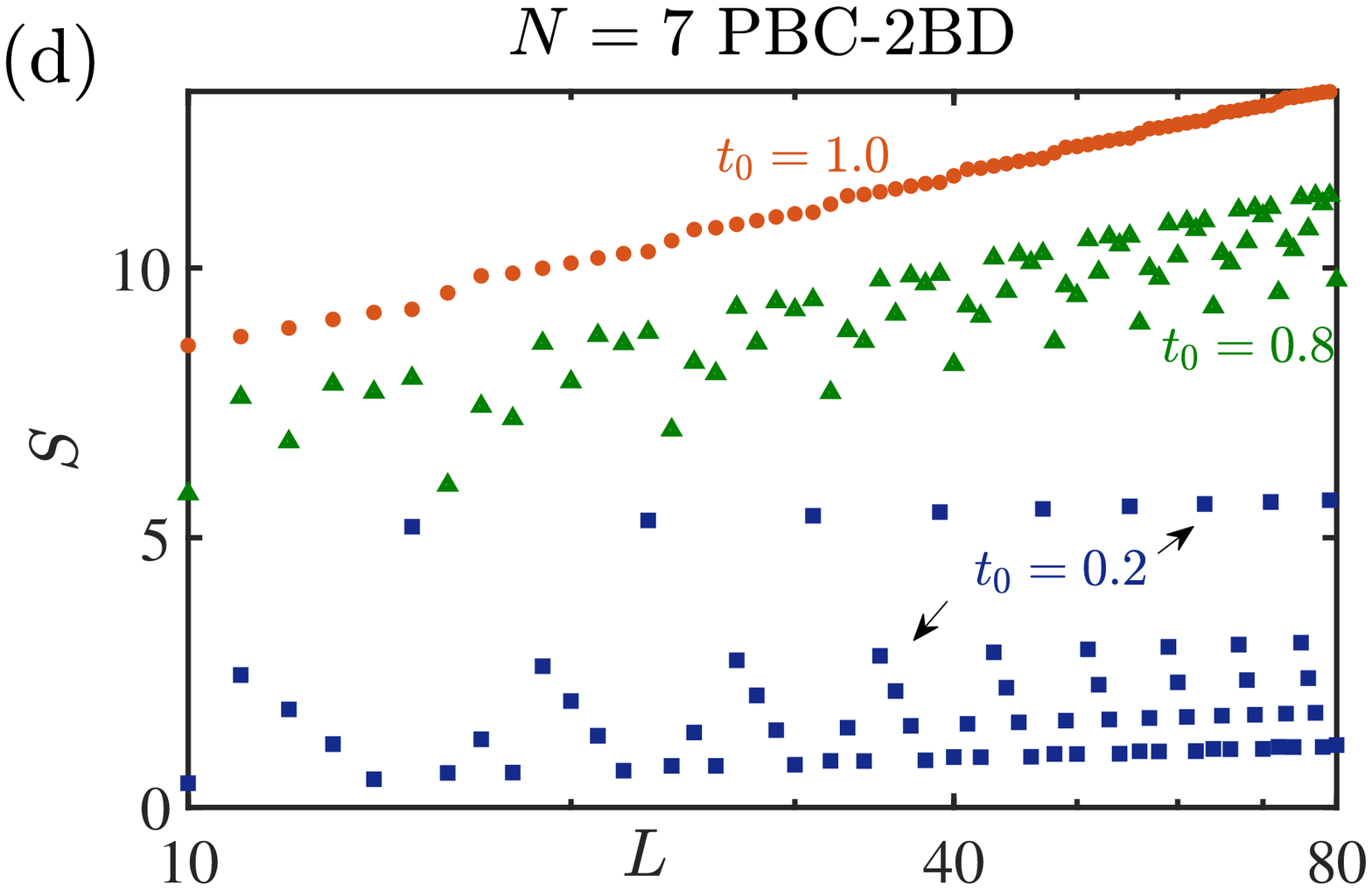}
\par\end{centering}
\caption{$S$ as a function of $L$ (log scale) for different values of $t_{0}$.
We have used here $t_{y}=t_{x}$, $\gamma_{n}=0$ and have considered
PBC and two bound defects, i.e., $t_{0}=t_{P}$.\label{fig:S_gamma=00003D0-PBC-2BD}}
\end{figure}
Let us now study the case $t_{P}=t_{0}$. This case can be understood
as ladders with PBC along the legs and 2 bound defects (2BD) at both
boundaries. We show the behavior of the von Neumann entropy, for ladder
with $N=2,3,4$ and $7$ legs, as a function of $L$ in Fig. \ref{fig:S_gamma=00003D0-PBC-2BD}.
For a given $N$, oscillations are known to be among sizes $L=m,m+N,m+2N,m+3N,\cdots$
for $m=1,2,3,\cdots$, for the homogeneous PBC ($t_{0}=1$). On the
other hand, likewise in the OBC situation, for non-homogeneous systems,
the sizes $L=m,m+(N+1),m+2(N+1),m+3(N+1),\cdots$, with $m=1,2,3,\cdots$,
will follow the same logarithmic branch behavior. Once again the ECC
was evaluated by fitting the data from each $S(L)$ branch to Eq.
(\ref{eq:S=00003DlogL}) and then taking the mean value among them.
It is worthy to stress here that for a PBC, there are two points of
contact between the subsystems and we need to set $\eta=2$ in Eq.
(\ref{eq:S=00003DlogL}). Taking into account our numerical precision,
we have found that $c_{\mathrm{eff}}$ has the same behavior presented
in Fig. \ref{fig:ceff_vs_N}(a) for all values of $t_{0}$ and number
of legs $N$.

\begin{figure}
\begin{centering}
\includegraphics[width=0.5\textwidth]{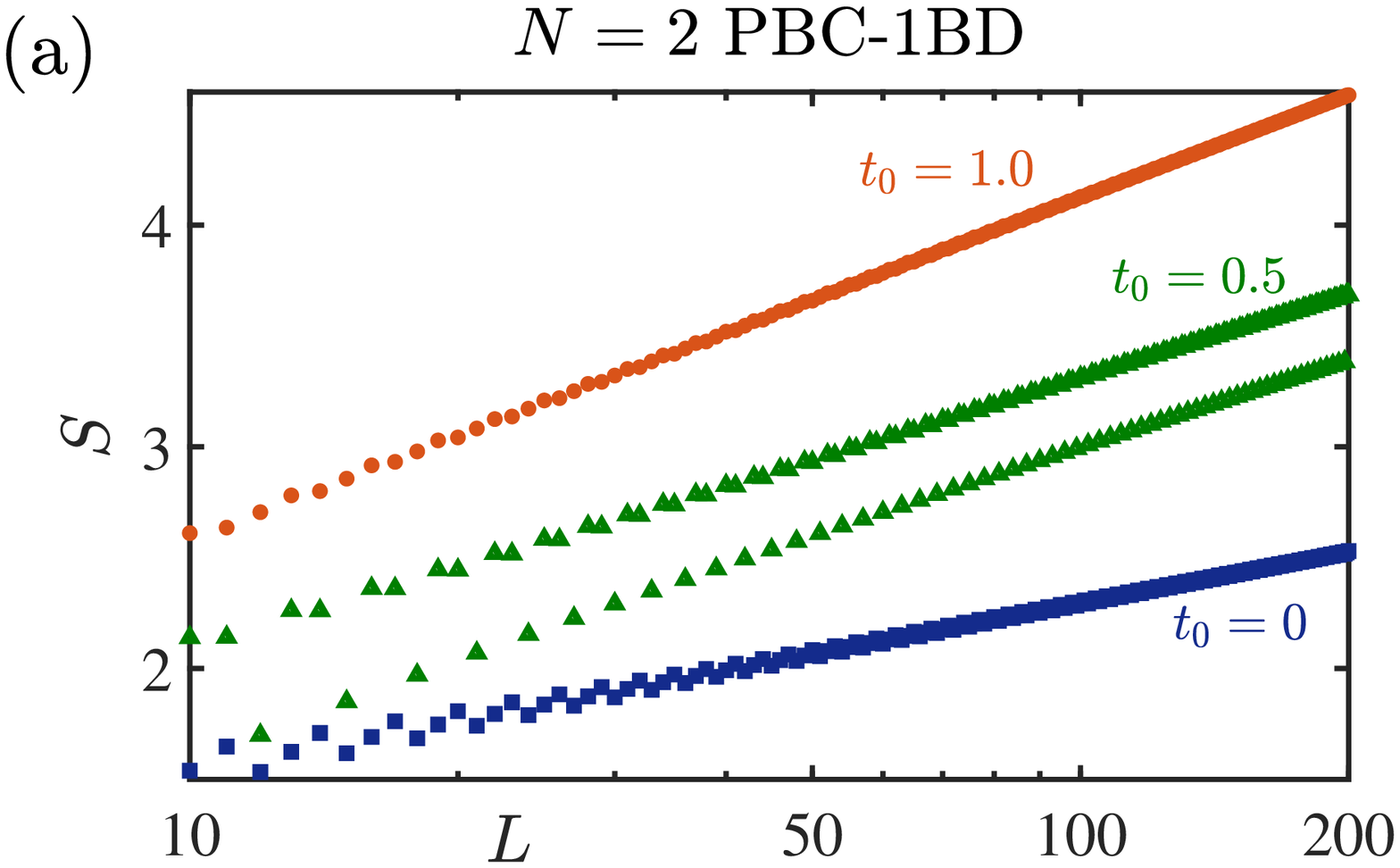}\includegraphics[width=0.5\textwidth]{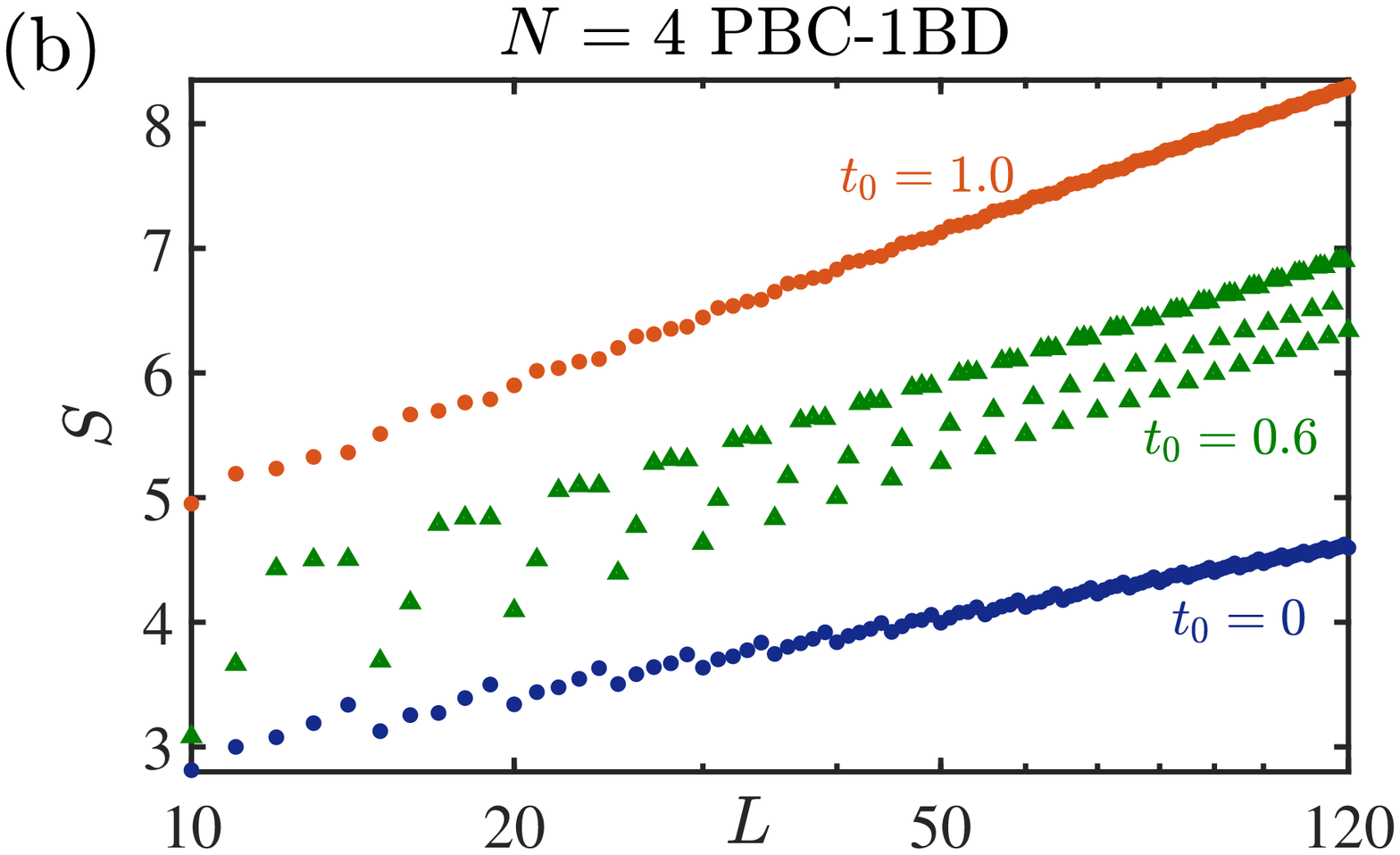}
\par\end{centering}
\begin{centering}
\includegraphics[width=0.5\textwidth]{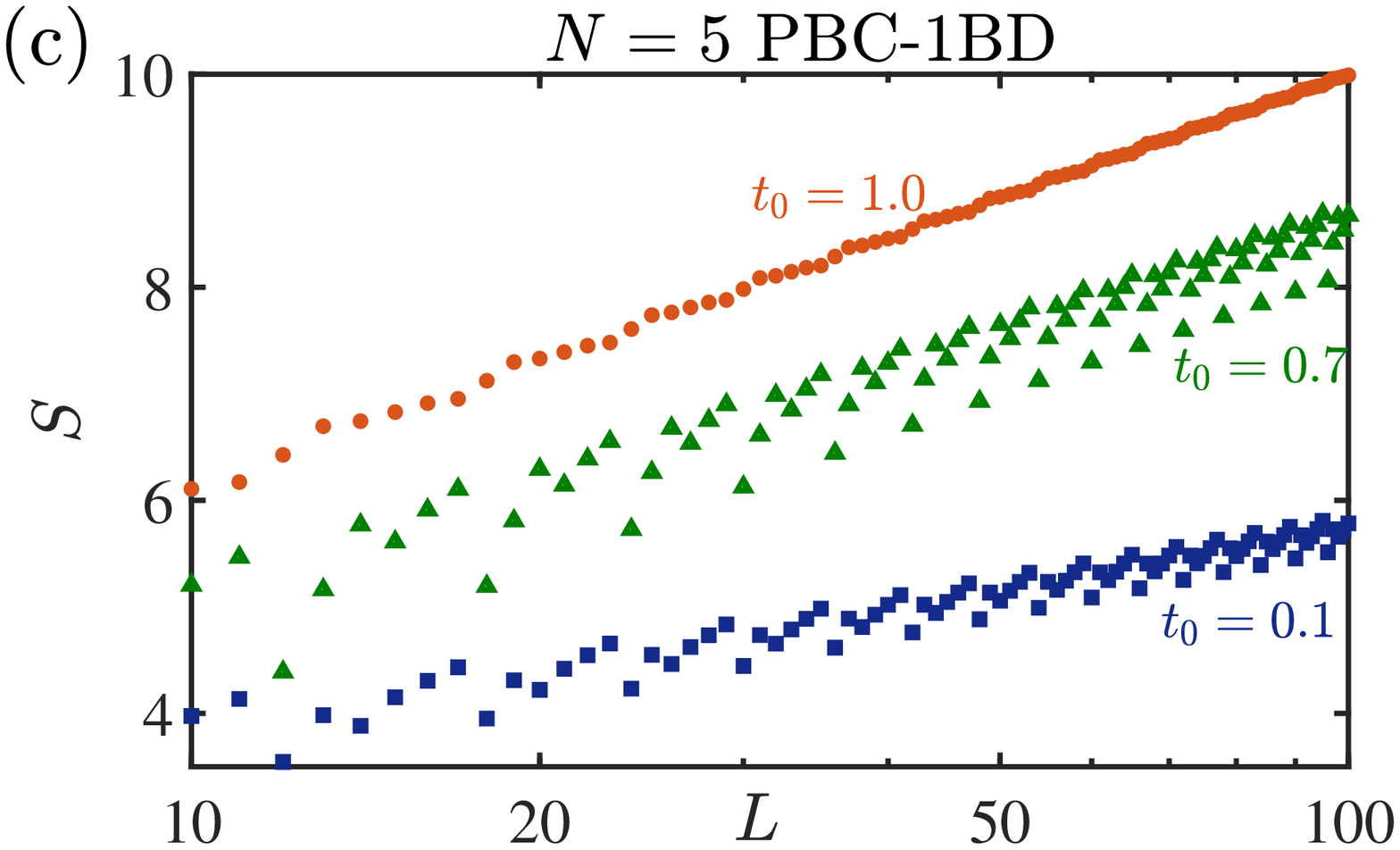}\includegraphics[width=0.5\textwidth]{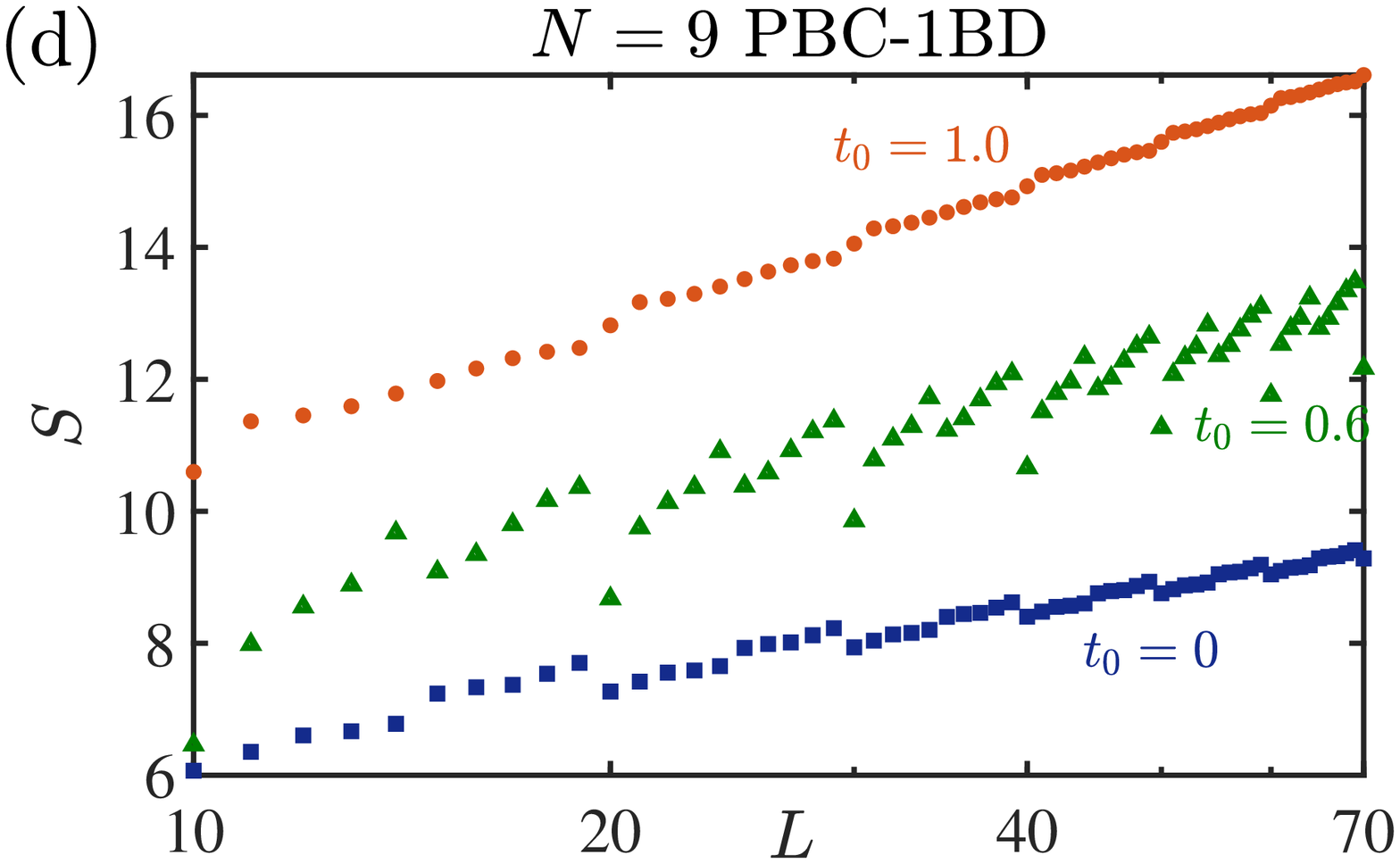}
\par\end{centering}
\caption{$S$ vs. $L$ (log scale) for different values of $t_{0}$ for (a)
$N=2$, (b) $N=4$, (c) $N=5$, and (d) $N=9$. We have set $t_{y}=t_{x}=t_{P}=1$
and $\gamma_{n}=0$.\label{fig:S_gamma=00003D0-PBC-1BD}}
\end{figure}
Last but not least we also show some results for the case $t_{P}=1$.
We call this situation as PBC with 1BC. Here, the case $t_{0}=1$
is again the homogeneous PBC (with no bound defect) ladder. On the
other hand, the case with $t_{0}=0$ is a homogeneous ladder with
OBC, in other words, it coincides exactly with the situation discussed
for the OBC system with $t_{0}=1$, the difference is that, here,
the boundary between the two subsystems is at the rungs $L$ and $-L+1$
and not at the rungs $n=0$ and $n=1$ as have used in Fig. \ref{fig:S_gamma=00003D0-OBC}.
We depict results of the quantum entanglement $S$ between the two
subsystems as a function of the these subsystems size $L$ in the
panels (a), (b), (c), and (d) of the Fig. \ref{fig:S_gamma=00003D0-PBC-1BD},
for two-leg ladder ($N=2$), four-leg ladder ($N=4$), five-leg ladder
($N=5$), and nine-leg ladder ($N=9$) systems, respectively. The
oscillation pattern follows the same branch rule of the PBC with 2BD
that we have just discussed. In order to obtain the ECC from Eq. (\ref{eq:S=00003DlogL})
we again need to use $\eta=2$ when we fit our data.

The results that we found for the central charge, as a function of
the defect amplitude $t_{0}$, are the following: at $t_{0}=1$, the
ECC is $c_{\mathrm{eff}}\approx N$, as expected, because the energy
dispersion has $N$ gapless modes; moreover, $c_{\mathrm{eff}}$ smoothly
decreased as $t_{0}$ decreases and it reaches $c_{\mathrm{eff}}\approx N/2$
when $t_{0}=0$. But, the behavior of $c_{\mathrm{eff}}$ for PBC
with one and two {[}the same of Fig. \ref{fig:ceff_vs_N}(a){]} boundary
defects are not independent of each other. We have also calculated
$c_{\mathrm{eff}}$ for 1BD for $N=1$ to $N=9$, and in fact, within
our numerical precision we were able to conclude that for PBC the
$c_{\mathrm{eff}}$ value for 2BD is double the $c_{\mathrm{eff}}$
for 1BD minus $N$. So, we can write the relations among them, for
the three different boundary conditions analyzed, as

\begin{equation}
c_{\mathrm{eff}}^{(\mathrm{PBC-2BD})}=2c_{\mathrm{eff}}^{(\mathrm{PBC-1BD})}-N=c_{\mathrm{eff}}^{(\mathrm{OBC})}.\label{eq:relacao_ceff}
\end{equation}

Another feature that we could conclude from this work is that there
is also a relationship between the ECC of the cases $\gamma=0$ and
$\gamma\neq0$. For $N=1$, it is well known that central charges
of the homogeneous cases ($|t_{0}|=1$ and/or $|t_{P}|=1$) are related
by $c(\gamma=0)=2\,c(\gamma\neq0)=1$. Furthermore, we found that
this aspect apply for the effective central charge for all $t_{0}$.
In other word, the non-homogeneous chain in absence of superconductivity
($\gamma=0$) has a $c_{\mathrm{eff}}(t_{0})$ that is double the
$c_{\mathrm{eff}}(t_{0})$ of the bi-quadratic non-homogeneous chain
with $\gamma\neq0$.

In this section we have only showed the behavior of $c_{\mathrm{eff}}(t_{0})$
in the range $0\leqslant t_{0}\leqslant1$. The reason for that is
because, after conducting extensive testing, we have notice that $c_{\mathrm{eff}}$
has the following symmetries $c_{\mathrm{eff}}(-t_{0})=c_{\mathrm{eff}}(t_{0})$
and $c_{\mathrm{eff}}(t_{0}^{-1})=c_{\mathrm{eff}}(t_{0})$.

\section{\label{sec:Conclusions}Conclusions}

The study of entanglement entropy in the many-body ground states of
hamiltonians composed of two halves can provide dramatic signatures
at a topological phase transition. We showed this by considering a
spinless fermionic toy model with multi-leg ladder geometry. Usual
correlation matrix techniques allowed us to compute the von Neumann
entanglement entropy.\cite{PESCHEL_JPA} The $N$-leg ladders we
investigated have rectangular geometries and allow hoppings $t_{y}$
along the rungs. On the other hand, along the legs, besides the nearest
neighbor hopping amplitude $t_{x}$, it is also considered a superconducting
pairing potential $\gamma$. Different boundary conditions were analyzed
along the legs and OBC were always fixed on the other direction. Our
work was divided in two cases $\gamma=0$ and $\gamma\neq0$.

From the presented study, one can claim that for a $N$-leg ladder,
when the superconducting gap $\gamma$ is nonzero, the quantum phase
diagrams show $2N$ critical lines {[}See Fig. \ref{fig:critical-mu}
and Eqs. (\ref{eq:q-valores}) and (\ref{fig:critical-mu}){]}. Moreover,
for the trivial case of non inter-leg hopping $t_{y}=0$, one have,
from Eq. (\ref{eq:mu_cr}), that $\mu_{\mathrm{cr}}=\pm2$, so there
are $N$ critical lines meeting at $(t_{y},\mu_{\mathrm{cr}})=(0,\pm2)$
on the phase diagram. Nevertheless, all the other possible interceptions
occur only between two critical lines. We also investigate the critical
points and lines using the ground state entanglement entropies between
the left and right sides of the ladders. In particular, we fixed $t_{y}$
($\mu$) and calculated $S$ as functions of $\mu$ ($t_{y}$). At
the critical lines, as a result of the divergence of the correlation
length, we also observed accentuated peaks on the von Neumann entropy
profile. Moreover, if the set of parameters $(t_{y},\mu_{\mathrm{cr}})$
coincides with an interception of two critical lines, the peak is,
in general, higher than the peaks where $(t_{y},\mu_{\mathrm{cr}})$
corresponds to single critical line without interception. In fact,
logarithmic divergence of entanglement entropy (as function of size
of a subsystem) is a signature of criticality in quantum models, in
particular, it is well know that for the simple case of a single leg,
the system is equivalent to the critical Ising chain in a transverse
magnetic field, in this case the central charge of the system is $c=1/2$.
Within our numerical precision, for $N>1$, we obtained that the effective
central change is $c_{\mathrm{eff}}=n_{\mathrm{GL}}c$, where $n_{\mathrm{GL}}$
is the number of gapless modes on the energy dispersion (which is
also exactly equal to the number of intercepting lines on the phase
diagram).

Our \textit{main results} are the entanglement properties that were
calculated when \textit{defects}, located at the contact points between
the subsystems are present. For both $\gamma=0$ and $\gamma\neq0$,
we have found that a logarithmic dependence of the von Neumann entropy
on the size persists, on the other hand, the amplitude (proportional
to the effective central charge) changes and becomes dependent on
the defect strength. Moreover, our study shows that the ECC continuously
decreases as the interface defect is turned on. In other words, for
small values of the interface hopping amplitudes, $t_{0}$ and/or
$t_{P}$, there are weak quantum correlations between the subsystems.
Analogously, when the interface hopping terms are strong ($t_{0},t_{P}\gg t_{x}=1$)
the subsystems should also be barely bipartite entangled with each
other.

Nevertheless, the ECC for $N>1$ are different for $\gamma=0$ and
$\gamma\neq0$. If the superconducting gap is non-null, the effective
central change is $c_{\mathrm{eff}}(N)=n_{\mathrm{GL}}c_{\mathrm{eff}}(1)$
even in the presence of defects. For the sake of example, regardless
of the number of legs $N$, when $n_{\mathrm{GL}}=1$, $c_{\mathrm{eff}}(N)$
has the same behavior of the ECC of a chain ($N=1$), additionally,
when two critical chemical potential lines intercept each other in
the phase diagram ($n_{\mathrm{GL}}=2$) $c_{\mathrm{eff}}(N)$ is
exactly twice the one of a single critical leg --- this happened
for all defect strength considered. On the other hand, in the absence
of a finite superconducting gap ($\gamma=0$), the number of gapless
modes $n_{\mathrm{GL}}$ is up to the number of legs $N$. In particular,
for $\mu=0$ we have that $n_{\mathrm{GL}}=N$. Another interesting
feature of the entanglement entropy in the region $\gamma=0$ is that
oscillations among different system-size parity appear naturally.
For the homogeneous case this oscillations were investigated in Ref.
\cite{Xavier_2014}. Moreover, for non-homogeneous, our results showed
that, in this case, $c_{\mathrm{eff}}(N)\leqslant n_{\mathrm{GL}}c_{\mathrm{eff}}(1)$
and the equal sign only holds when the system is homogeneous across
the interface between the two subsystems. That is to say in the presence
of interface defects the ECC of ladders with inter-leg hopping becomes
smaller than the one in the absence of hopping mechanism among the
rungs. For example, two disconnected chains have an ECC that is double
the one of a single chain, nevertheless, if the two chains are connected
to one another through inter-leg hopping the ECC becomes smaller than
$2c_{\mathrm{eff}}(1)$ in the presence of a defect. 

This last result shows that for $\gamma=0$ the entanglement entropy
between the left and right sides of the system depends on the (ratio
of the) hopping strengths along legs and rungs in the presence of
defects. Moreover, one can expected that the entanglement should decrease
for all $t_{y}\neq t_{x}$, i.e., we speculate that the presence of
the hoppings along the rungs makes them correlated and this will make
the quantum correlations along the legs weaker. Further investigations
on that could provide a finer understanding of quasi-one-dimensional
non-homogeneous systems. We believe that our work also opens up room
for further exploration of how the addiction of superconducting pairing
terms along the rungs of the ladder would change the phase diagrams
and the entanglement properties.

\end{document}